\documentclass[article,twocolumn]{revtex4}

\usepackage{graphicx}
\usepackage{epsfig}
\usepackage{xspace}
\usepackage{amsmath}
\usepackage{float}
\DeclareMathSymbol{\Delta}{\mathalpha}{letters}{"01}
\DeclareMathSymbol{\Sigma}{\mathalpha}{letters}{"06}
\DeclareMathSymbol{\Lambda}{\mathalpha}{letters}{"03}
\DeclareMathSymbol{\Xi}{\mathalpha}{letters}{"04}

\newcommand{\rh}{\ensuremath{\rho(770)^{0}}\xspace}
\newcommand{\ks}{\ensuremath{K^{*}(892)^{0}}\xspace}
\newcommand{\aks}{\ensuremath{\overline{K}^{*}(892)^{0}}\xspace}
\newcommand{\ph}{\ensuremath{\phi(1020)}\xspace}
\newcommand{\dl}{\ensuremath{\Delta(1232)^{++}}\xspace}

\newcommand{\ssx}{\ensuremath{\Sigma(1385)^{\pm}}\xspace}
\newcommand{\assx}{\ensuremath{\overline{\Sigma}(1385)^{\mp}}\xspace}
\newcommand{\ls}{\ensuremath{\Lambda(1520)}\xspace}
\newcommand{\als}{\ensuremath{\overline{\Lambda}(1520)}\xspace}
\newcommand{\xs}{\ensuremath{\Xi(1530)^{0}}\xspace}
\newcommand{\axs}{\ensuremath{\overline{\Xi}(1530)^{0}}\xspace}

\newcommand{\rhpi}{\ensuremath{\rh/\pi^{\pm}}\xspace}

\newcommand{\ksk}{\ensuremath{\ks/K}\xspace}

\newcommand{\phk}{\ensuremath{\ph/K}\xspace}
\newcommand{\dlp}{\ensuremath{\dl/p}\xspace}
\newcommand{\ssl}{\ensuremath{\ssx/\Lambda}\xspace}
\newcommand{\lsl}{\ensuremath{\ls/\Lambda}\xspace}
\newcommand{\xsx}{\ensuremath{\xs/\Xi}\xspace}

\newcommand{\pb}{\mbox{Pb-Pb}\xspace}
\newcommand{\pp}{\mbox{$p$-$p$}\xspace}
\newcommand{\ppb}{\mbox{$p$-Pb}\xspace}
\newcommand{\ada}{\mbox{$A$-$A$}\xspace}

\newcommand{\rs}[1][7~TeV]{\ensuremath{\sqrt{s}=}~#1\xspace}

\newcommand{\rsnn}[1][2.76~TeV]{\ensuremath{\sqrt{s_{NN}}=}~#1\xspace}
\newcommand{\rsnnppb}[1][5.02~TeV]{\ensuremath{\sqrt{s_{NN}}=}~#1\xspace}

\newcommand{\raa}{$R_{AA}$\xspace}

\newcommand{\dnc}{\ensuremath{\langle dN_{\mathrm{ch}}\kern-0.06em /\kern-0.13em d\eta\rangle}\xspace}
\newcommand{\dncr}{\ensuremath{\dnc^{1/3}}\xspace}

\newcommand{\pT}{\ensuremath{p_{\mathrm{T}}}\xspace}
\newcommand{\mpt}{\ensuremath{\langle\pT\rangle}\xspace}
\newcommand{\gvc}{\ensuremath{\mathrm{GeV}/c}\xspace}

\begin{document}

\hyphenpenalty=100000

\title[]{Hadronic resonance production and interaction in \ppb collisions at LHC energies in EPOS3}

%
\author{A. G. Knospe$^{1\dagger}$, C. Markert$^2$, K. Werner$^3$, J. Steinheimer$^4$, M. Bleicher$^{5,6}$}

\affiliation{
$^1$ Lehigh University, Department of Physics, 16 Memorial Drive East, Bethlehem, Pennsylvania 18015, USA \\
$^{\dagger}$ formerly with the University of Houston, Department of Physics, 3507 Cullen Blvd., Houston, Texas 77204-5005, USA \\
$^2$ The University of Texas at Austin, Department of Physics, 2515 Speedway, C1600, Austin, Texas 78712-0264, USA \\
$^3$ SUBATECH, UMR 6457, Universit$\acute{e}$ de Nantes, Ecole des Mines de Nantes, IN2P3/CNRS. 4 rue Alfred Kastler, 44307 Nantes CEDEX 3, France \\
$^4$Frankfurt Institute for Advanced Studies, Ruth-Moufang-Stra{\ss}e 1, 60438 Frankfurt am Main, Germany\\
$^5$ Institut f\"ur Theoretische Physik, Johann Wolfgang Goethe-Universit\"at Frankfurt am Main,
Max-von-Laue-Stra{\ss}e 1, 60438 Frankfurt am Main, Germany\\
$^6$Helmholtz Research Academy Hesse for FAIR (HFHF), GSI Helmholtz Center, Campus Frankfurt, Max-von-Laue-Stra{\ss}e 12, 60438 Frankfurt am Main, Germany}

\vspace*{1cm}

\begin{abstract}

Using the EPOS3 model with UrQMD to describe the hadronic phase, we study the production of short-lived hadronic resonances and the modification of their yields and \pT spectra in \ppb collisions at \rsnnppb. High-multiplicity \ppb collisions exhibit similar behavior to mid-peripheral \pb collisions at LHC energies, and we find indications of a short-lived hadronic phase in \ppb collisions that can modify resonance yields and \pT spectra through scattering processes. The evolution of resonance production is investigated as a function of the system size, which is related to the lifetime of the hadronic phase, in order to study the onset of collective effects in \ppb collisions. We also study hadron production separately in the core and corona parts of these collisions, and explore how this division affects the total particle yields as the system size increases.

\end{abstract}


\maketitle

\section{Introduction}

In heavy-ion collisions at LHC energies, the Quark Gluon Plasma (QGP), a state of partonic matter consisting of deconfined quarks and gluons, is expected to be created. As the system expands and cools, it undergoes a phase transition into a gas of hadrons. Inelastic interactions among the hadrons stop at the ``chemical freeze out" at temperature $T_{\rm ch}$ and elastic interactions stop at the ``kinetic freeze out" at temperature $T_{\rm kin}$. Hadronic resonances with lifetimes on the order of a few fm/$c$ are sensitive probes of the ``fireball" due to the fact that they can decay and re-form throughout the full evolution of the hadronic phase~\cite{Markert_Vitev}. Even after chemical freeze out, long-lived particles may pseudo-elastically scatter through a resonance state, thus increasing the resonance yield. Conversely, if short-lived resonances decay during the hadronic phase, their decay products may scatter with other components of the hadron gas (either elastically or pseudo-elastically through a different resonance); such ``re-scattering" inhibits the reconstruction of the original resonance and reduces the measured yield. Measurements of resonances may therefore help us understand the properties of the hadronic phase, which influence the relative \pT-dependent strengths of the regeneration and re-scattering effects.

Recent measurements from the LHC experiments have shown that high-multiplicity \pp and \ppb collisions exhibit similar behavior to peripheral \pb collisions. The ``hadrochemistry" of the system (the abundances of different hadron species) depends primarily on the charged-particle multiplicity of the collision, \textit{i.e.}, for a given multiplicity, hadron abundances are the same regardless of the collision system~\cite{ALICE_strangeness_pp7,ALICE_LF_pp7}. As the multiplicity increases in \pp and \ppb collisions, hadron \pT spectra harden and the $p/\pi$ and $\Lambda/K^{0}_{\mathrm{S}}$ ratios are enhanced at intermediate \pT~\cite{ALICE_LF_pp7,ALICE_LF_pPb,ALICE_Kstar_phi_pPb}; qualitatively similar behavior is also seen in \pb collisions, where it may be attributable to collective flow~\cite{ALICE_piKp_highpT_PbPb,ALICE_K0S_Lambda_PbPb}. There are also hints of a multiplicity-dependent suppression of the yields of \ks mesons in \pp and \ppb collisions~\cite{ALICE_LF_pp7,ALICE_Kstar_phi_pPb}, qualitatively similar to what is observed in \pb collisions~\cite{ALICE_Kstar_phi_PbPb,ALICE_Kstar_phi_highpT_PbPb}. These results raise the question of whether collective effects may be present in the smaller collision systems.

In a previous paper~\cite{Knospe_EPOS_PbPb}, we studied the production of resonances in \pb collisions at \rsnn using the EPOS3 framework~\cite{epos1,epos2,epos3}, which includes the UrQMD model~\cite{Bass:1998ca,Bleicher:1999xi,Steinheimer:2017vju} for the description of hadronic interactions in the hadronic phase. This paper extends that study to \ppb collisions at \rsnnppb and presents additional results for the \pb collision system. We use hadronic resonances, specifically modifications of their yields and \pT spectra, to test the existence of a hadronic phase with non-zero lifetime in \ppb collisions. We have produced 1.8 million p-Pb collisions with UrQMD turned on and the same number with UrQMD turned off. We have followed the approach of the ALICE Collaboration~\cite{ALICE_LF_pPb} and divided the \ppb event sample into multiplicity classes using the charged-particle multiplicity in the pseudorapidity range $2.8<\eta_{\mathrm{lab}}<5.1$ (in the direction of the Pb beam), which is the same range spanned by the ALICE V0A scintillator array~\cite{ALICE_V0}.

\begin{figure}
\centering
\includegraphics[angle=0,scale=0.42]{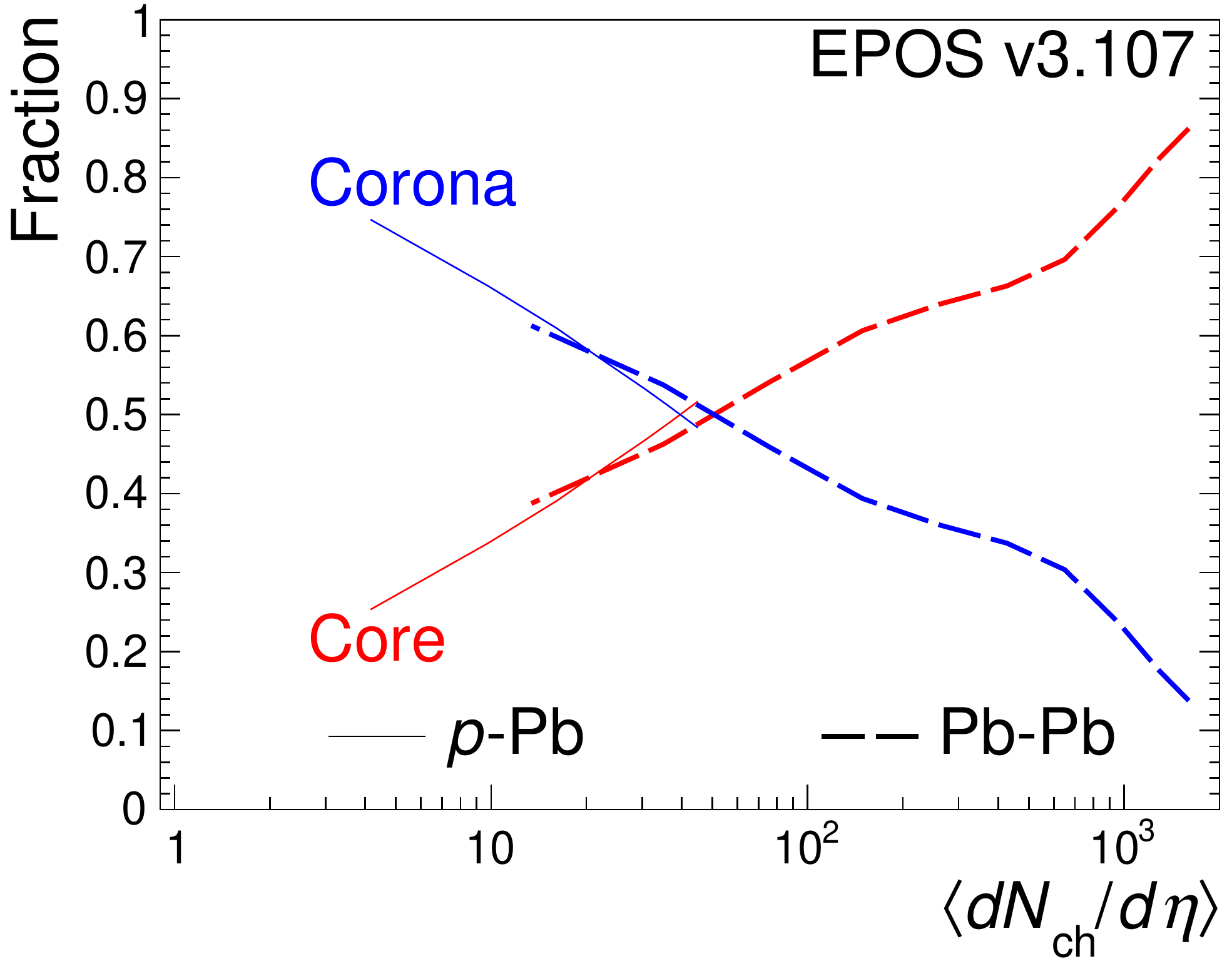}
\caption{\textbf{(a):} Fraction of particles originating from the core and corona parts of \ppb collisions at \rsnnppb and \pb collisions at \rsnn when UrQMD is turned off.}
\label{fig_coco_ratio}
\end{figure}

The EPOS3 model~\cite{epos1,epos2,epos3} describes the full evolution of a heavy-ion collision. The initial stage is treated via a multiple-scattering approach based on Pomerons and strings. The produced string segments are divided into core and corona contributions~\cite{epos4}. The core is taken as the initial condition for QGP evolution, for which we employ 3+1D viscous hydrodynamics. The corona part is simply composed of hadrons from string decays. The division between the core and corona parts of the collision is illustrated in Fig.~\ref{fig_coco_ratio}, which shows the fractions of long-lived hadrons ($\pi$, K, p, $\Lambda$, $\Sigma$, $\Xi$, $\Omega$, and their antiparticles) produced in the different parts of the collisions, for both \ppb collisions at \rsnnppb and the \pb collision at \rsnn used in our previous paper~\cite{Knospe_EPOS_PbPb}. Low-multiplicity (peripheral) collisions are dominated the corona, while central \pb collisions are core-dominated. High-multiplicity \ppb collisions have approximately equal contributions from the core and the corona. After hadronization~\cite{cooperfrye} of the fluid (core part), all hadrons, including those from the corona, are fed into UrQMD~\cite{Bass:1998ca,Bleicher:1999xi,Steinheimer:2017vju}, which describes hadronic interactions in a microscopic approach. The chemical and kinetic freeze outs occur within this phase. Resonance signals have been previously studied using the UrQMD model~\cite{urqmd_high_pt,urqmd_sis,urqmd_2002,urqmd_2003,urqmd_2004,urqmd_2005,urqmd_2006,urqmd_2008,Steinheimer:2015msa,Steinheimer2012,Steinheimer:2017vju}.

\begin{figure}
\centering
\includegraphics[angle=0,scale=0.42]{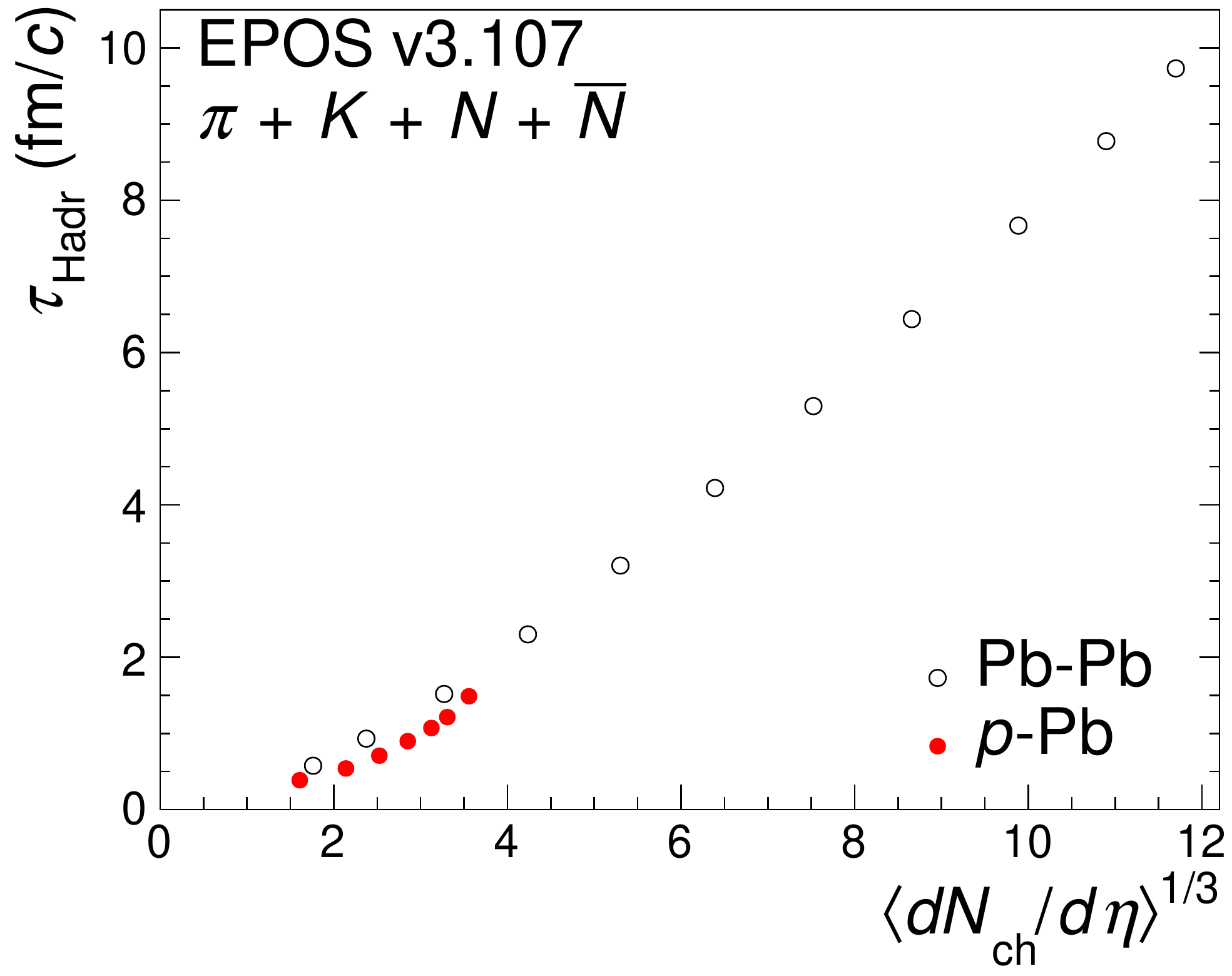}
\caption{Lifetime of hadronic phase in \ppb (red full circles) and \pb (black open circles)~\cite{Knospe_EPOS_PbPb} collisions.}
\label{fig_lifetime}
\end{figure}

For more central collisions, the hadronic phase lasts longer (\textit{i.e.}, the time between chemical and kinetic freeze out increases), which would imply more hadronic interactions and could result in greater modification of resonance \pT spectra and yields. The hadronic lifetime estimated from EPOS3 calculations increases from 0.5 to 10 fm/$c$ depending on centrality in \pb collisions as shown in Fig.~\ref{fig_lifetime}. Here, we use \dncr, the cube root of the mean charged-particle multiplicity measured by ALICE~\cite{ALICE_LF_pPb,ALICE_multiplicity_PbPb} at mid-rapidity ($|\eta_{\mathrm{lab}}|<0.5$), as a proxy for the system size. Our new estimates for the hadronic-phase lifetime in \ppb collisions appear to follow a similar trend to that seen in \pb collisions~\cite{Knospe_EPOS_PbPb}, with the highest multiplicity \ppb collisions (0-5\%) having similar hadronic lifetimes ($\approx1.5$ fm/$c$) and multiplicities to peripheral \pb collisions (70-80\%). It should be noted that while these collisions produce similar charged-hadron multiplicities, they have quite different geometries. The multiplicity of charged hadrons (mostly pions) scales with the number of participant nucleons, and therefore scales with the event activity.

\section{Resonance Reconstruction}

Experimentally, hadronic resonances are reconstructed using the invariant mass method via measurements of the momenta of their decay daughters. Charged pions, charged kaons, and (anti)protons are often identified through measurements of energy loss $(dE/dx)$ in a Time Projection Chamber (TPC) and/or the velocity in a Time-of-Flight (TOF) detector. Weakly decaying particles, such as $\Lambda$ and $\Xi$, can be selected based on their decay topologies, which adds further constraints. Table~\ref{decay}~\cite{PDG} lists the specific decay channels investigated in the EPOS3 approach, which are the same channels used experimentally by STAR and ALICE. In these model calculations, resonances that decay via the channels listed in Table~\ref{decay} are flagged and the decay products are followed throughout the system evolution. If neither decay product undergoes a re-scattering, the resonance is flagged as reconstructable. Throughout this paper, the resonance yields, both from these calculations and from experiment, are corrected by the appropriate branching ratio. The shorthand notations listed in the second column of Table~\ref{decay} will sometimes be used to denote these resonances. Results for particles and antiparticles are always combined, even when not explicitly noted.

\begin{table}[h!]
\centering
\caption{The resonances are constructed experimentally via the decay channels listed~\cite{PDG}. These same decays are used in our studies of resonances in EPOS3 and UrQMD.}
\label{table2}
\begin{ruledtabular}
\begin{tabular}{lcccc}
 & & Decay & Branching & Lifetime \\
Resonance & Shorthand & Channel & Ratio & (fm/$c$) \\
\hline
$\rho(770)^{0}$ & $\rho^{0}$ & $\pi^{+}$ + $\pi^{-}$  & 1 & 1.335 \\
$K^{*}(892)^{0}$ & $K^{*0}$ & $\pi^{-}$ + $K^{+}$ & 0.67 & 4.16 \\
$\phi(1020)$ & $\phi$ & $K^{+}$ + $K^{-}$ & 0.492 & 46.26 \\
$\Delta(1232)^{++}$ & $\Delta^{++}$ & $\pi^{+}$ + $p$ & 1 & 1.69 \\
$\Sigma(1385)^{+}$ & $\Sigma^{*+}$ & $\pi^{+}$ + $\Lambda$ & 0.870 & 5.48 \\
$\Sigma(1385)^{-}$ & $\Sigma^{*-}$ & $\pi^{-}$ + $\Lambda$ & 0.870 & 5.01 \\
$\Lambda(1520)$ & $\Lambda^{*}$ & $K^{-}$ + $p$ & 0.225 & 12.54\\
$\Xi(1530)^{0}$ & $\Xi^{*0}$ & $\pi^{+}$ + $\Xi^{-}$ & 0.67 & 22 \\
\end{tabular}
\end{ruledtabular}
\label{decay}
\end{table}

\begin{figure*}[!ht]
\centering
\includegraphics[width=30pc]{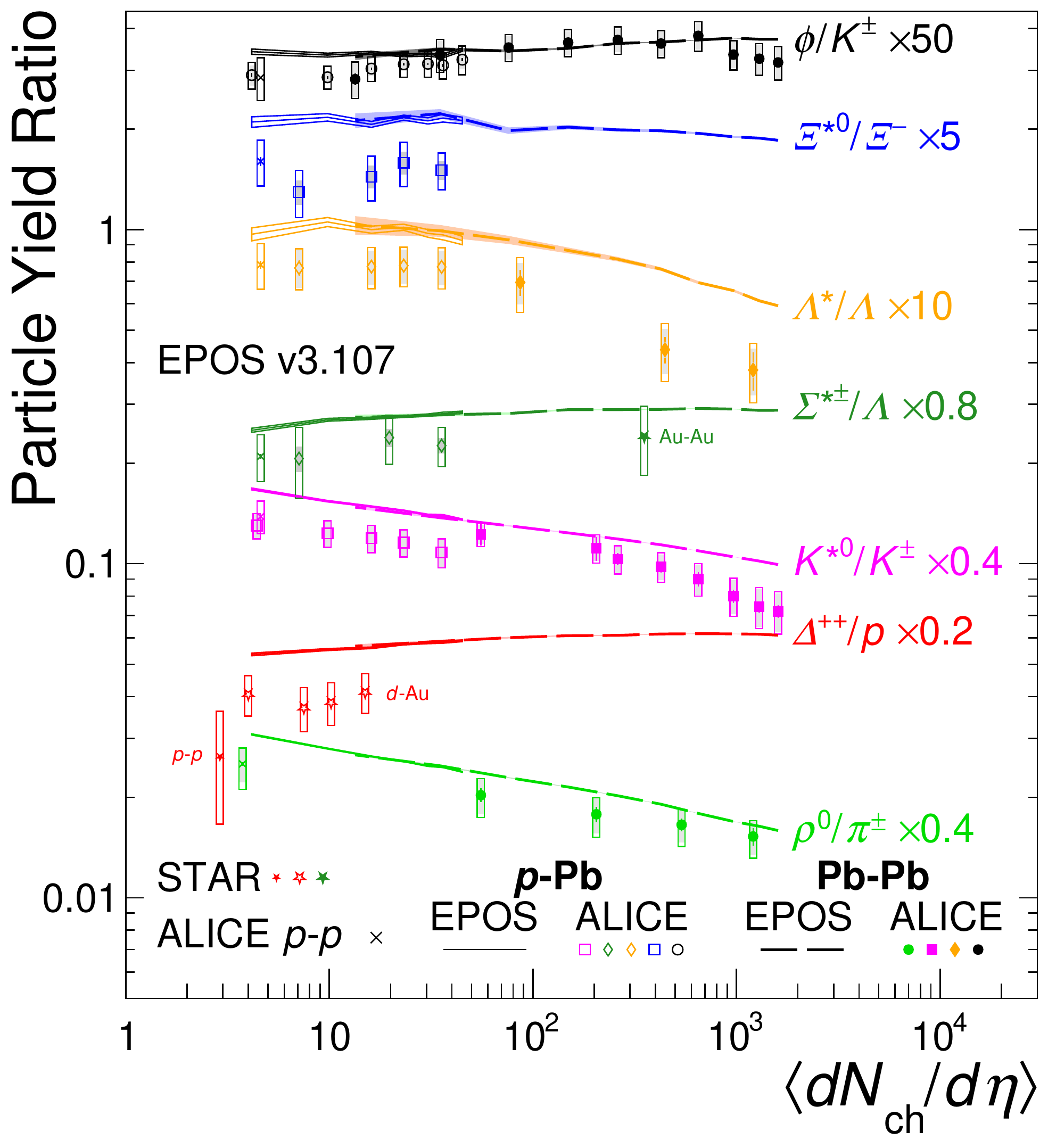}
\caption{Ratios of resonance yields to long-lived hadrons as functions of the charged-particle multiplicity measured at mid-rapidity. EPOS3 data are shown for \pb collisions at \rsnn~\cite{Knospe_EPOS_PbPb} (thick lines) and \ppb collisions at \rsnnppb (thin lines), with bands representing statistical uncertainties. The EPOS3 results are compared to experimental results, mostly from ALICE~\cite{ALICE_Kstar_phi_pPb,ALICE_Kstar_phi_PbPb,ALICE_Kstar_phi_highpT_PbPb,ALICE_Sigmastar_Xistar_pPb,ALICE_rho_PbPb,ALICE_Lambdastar_pPb,ALICE_Lambdastar_PbPb}. In a few cases, ALICE data are not available and STAR data for \rsnn[200~GeV] are shown instead~\cite{STAR_Sigmastar_Lambdastar_AuAu,STAR_resonances_dAu}. For the experimental data, bars represent statistical uncertainties, open boxes represent the total systematic uncertainties, and shaded boxes represent systematic uncertainties uncorrelated between multiplicity classes. The particle symbols denote both particles and antiparticles.}
\label{fig_ratios}
\end{figure*}

\section{Resonance Yields and Ratios}

Figure~\ref{fig_ratios} shows a summary of EPOS3 calculations of the ratios of resonance yields to those of long-lived hadrons (usually with the same strange-quark content) as a function of event multiplicity for \pb~\cite{Knospe_EPOS_PbPb} collisions, along with our new calculations for \ppb collisions. The abscissa \dnc is commonly used as a proxy for the event activity and is often used to compare results from different collision systems. Its use in nucleus-nucleus collisions is connected to femtoscopy studies~\cite{ALICE_HBT_2011,Lisa_FemtoscopyReview,Graef2012}, which suggest that it scales in proportion to the radius of the collision system. Under the simple assumption that the probability of re-scattering is proportional to the distance traveled through the hadronic medium, an exponential decrease in measured resonance yields as a function of the system radius or \dncr might be expected. The EPOS3 results are compared to experimental results from ALICE. In a few cases, ALICE measurements are unavailable and measurements from STAR are used instead. It should be noted that while the STAR results are from lower energies than the EPOS3 calculations, these ratios do not generally depend strongly on collision energy. It is notable that particle yield ratios calculated by EPOS3 in \ppb collisions are consistent with the \pb values for similar multiplicities, even though the initial geometries of the collision systems are very different.

\begin{figure*}[!ht]
\centering
\includegraphics[angle=0,scale=0.42]{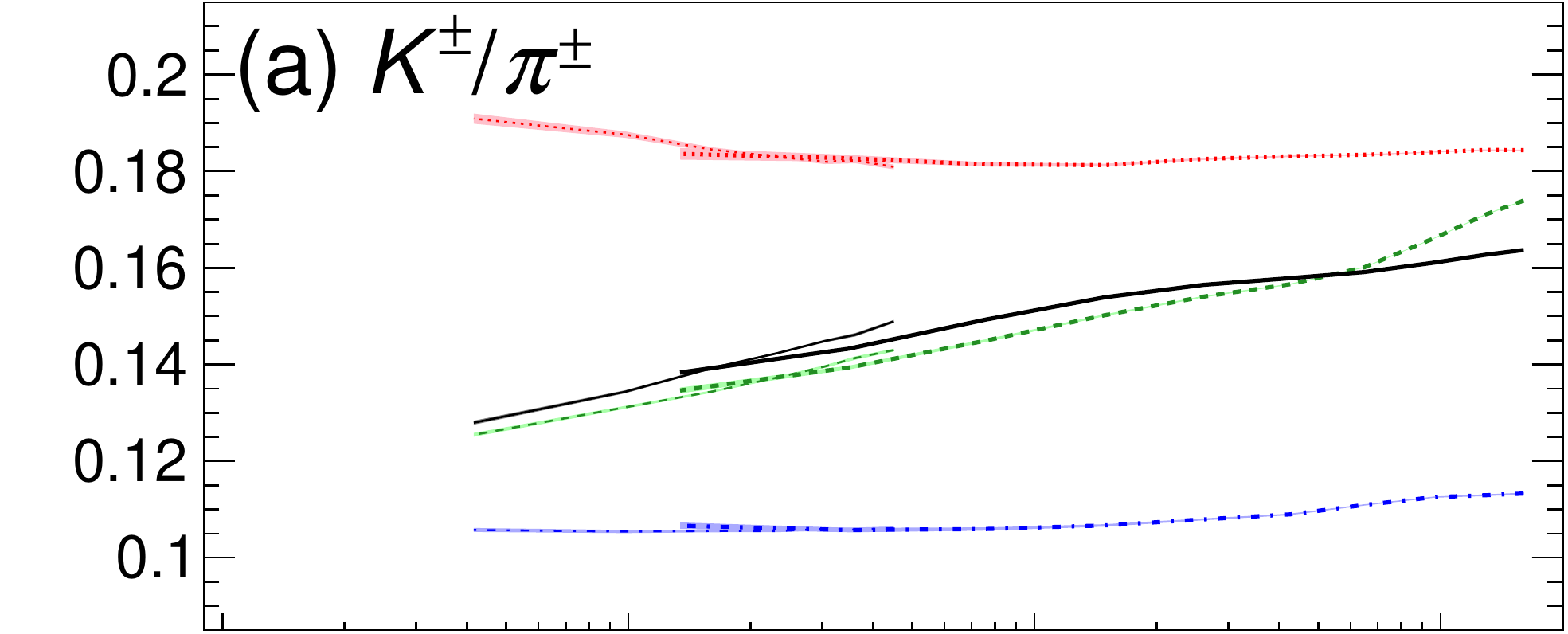}
\includegraphics[angle=0,scale=0.42]{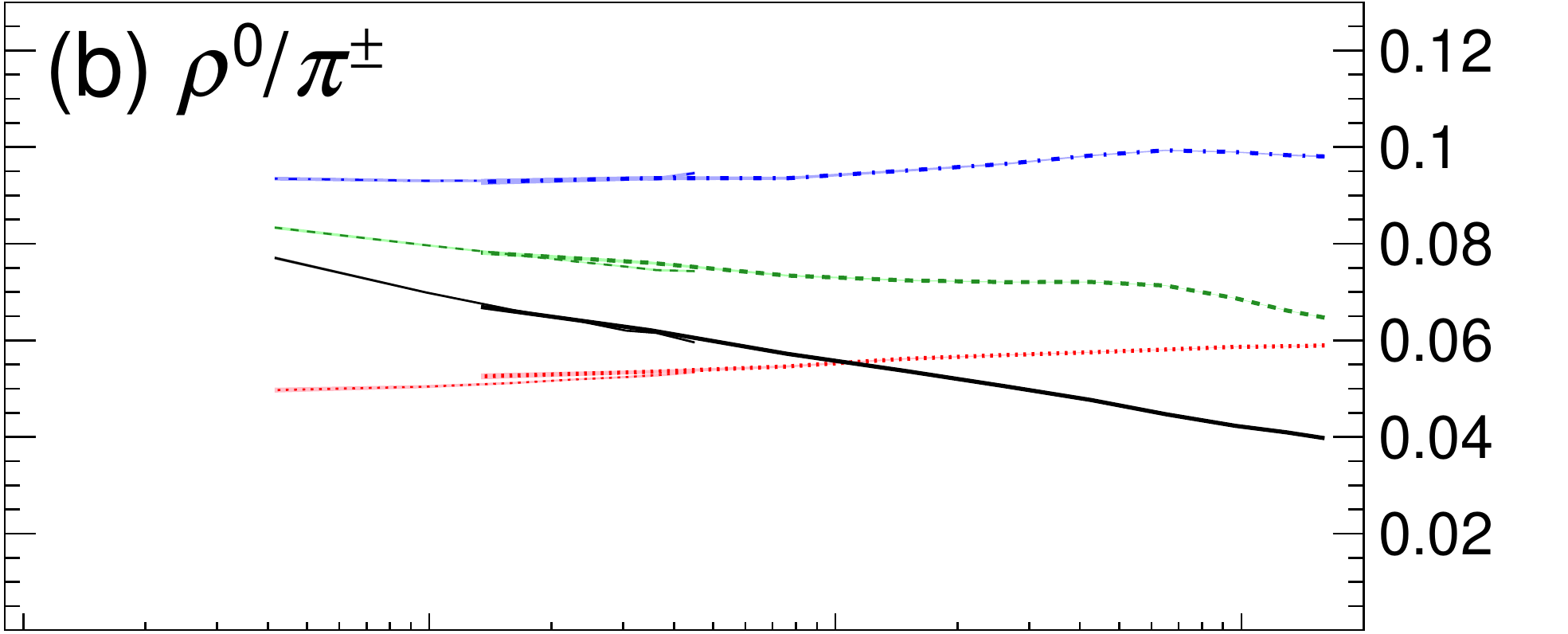}
\includegraphics[angle=0,scale=0.42]{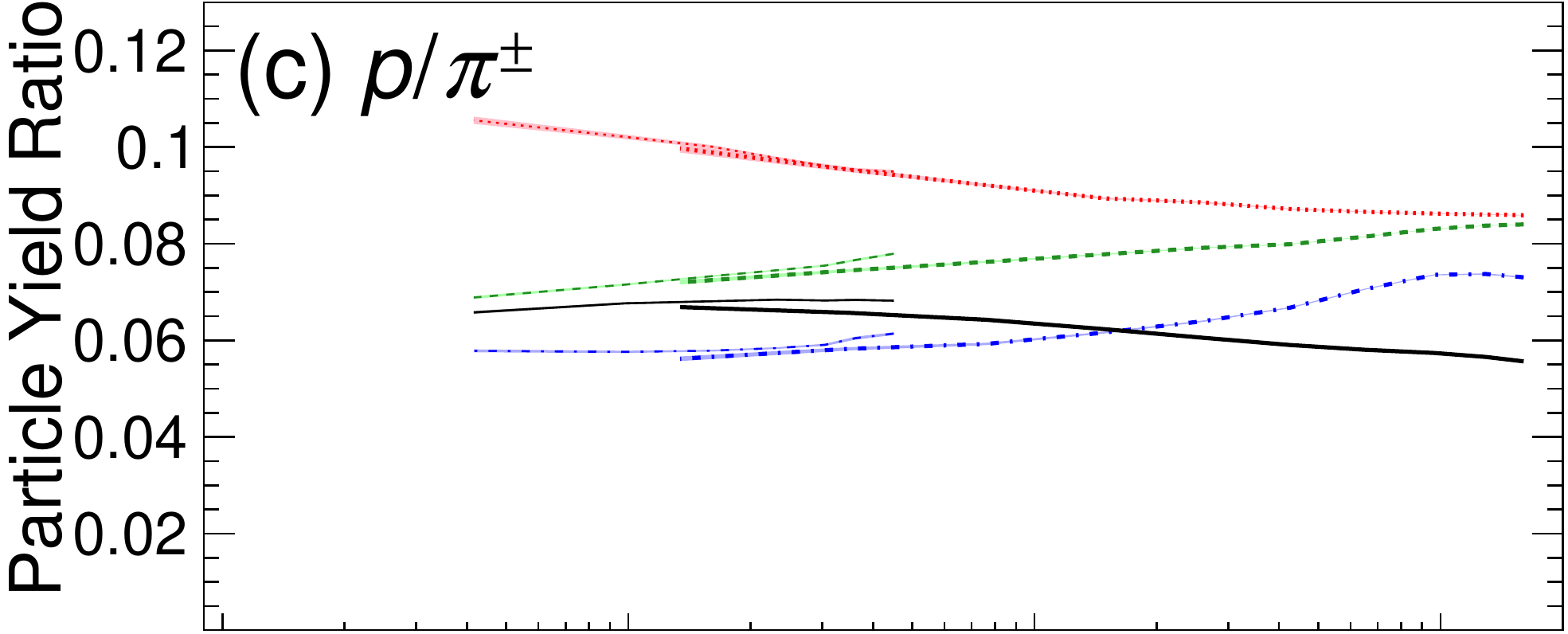}
\includegraphics[angle=0,scale=0.42]{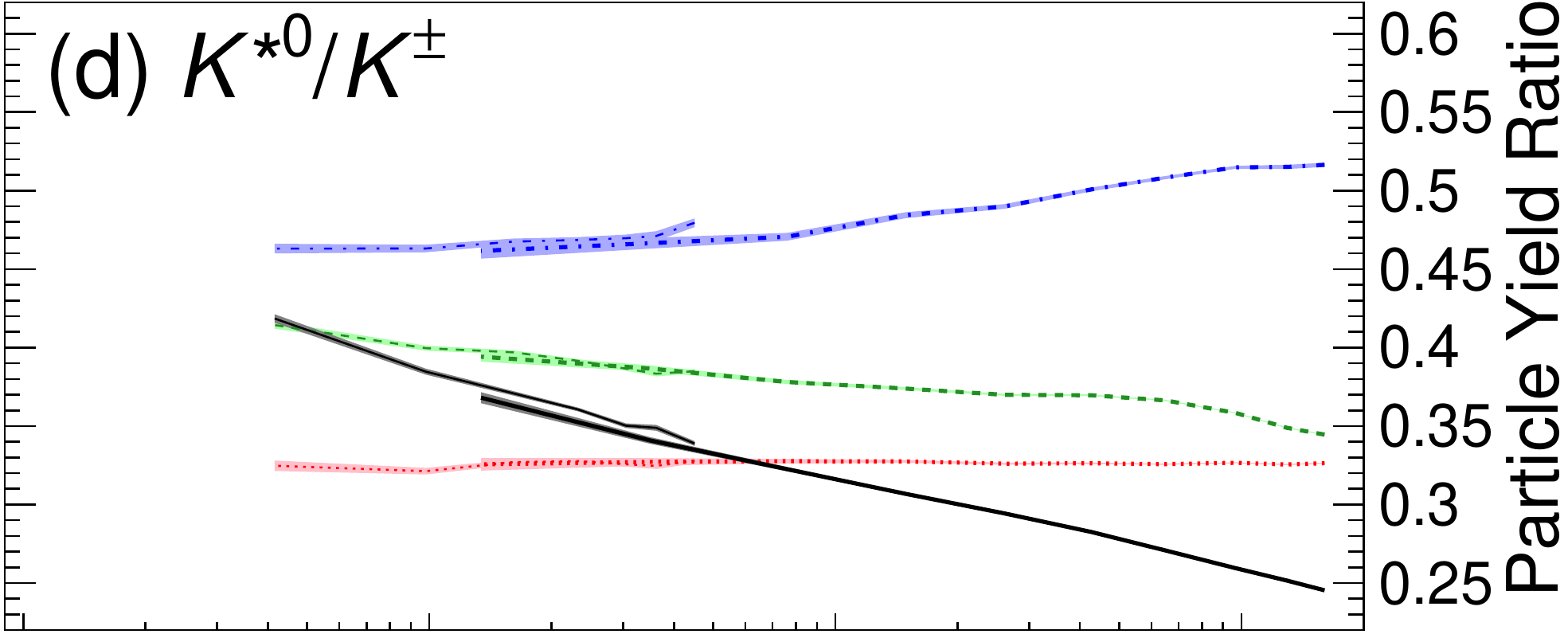}
\includegraphics[angle=0,scale=0.42]{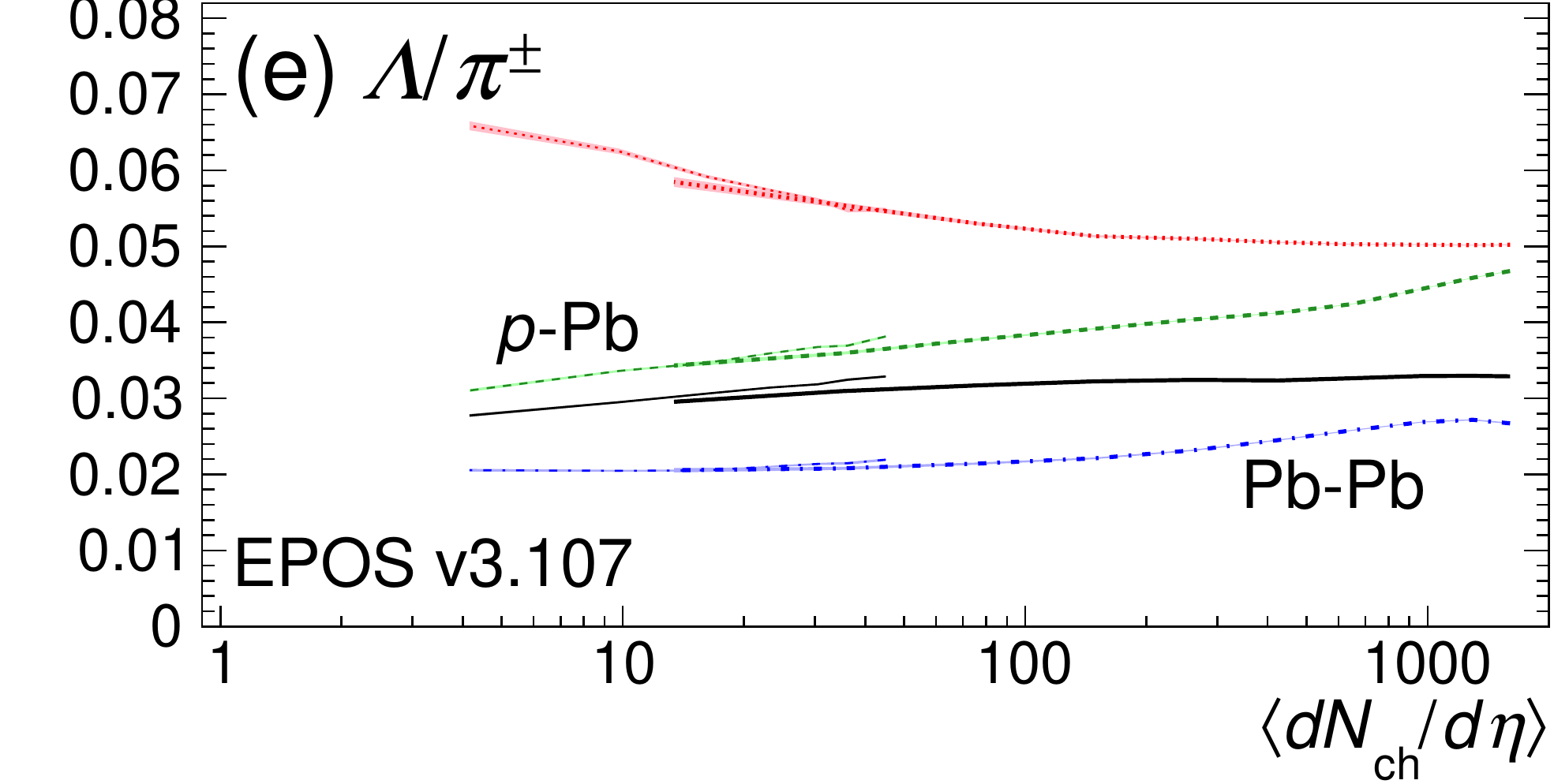}
\includegraphics[angle=0,scale=0.42]{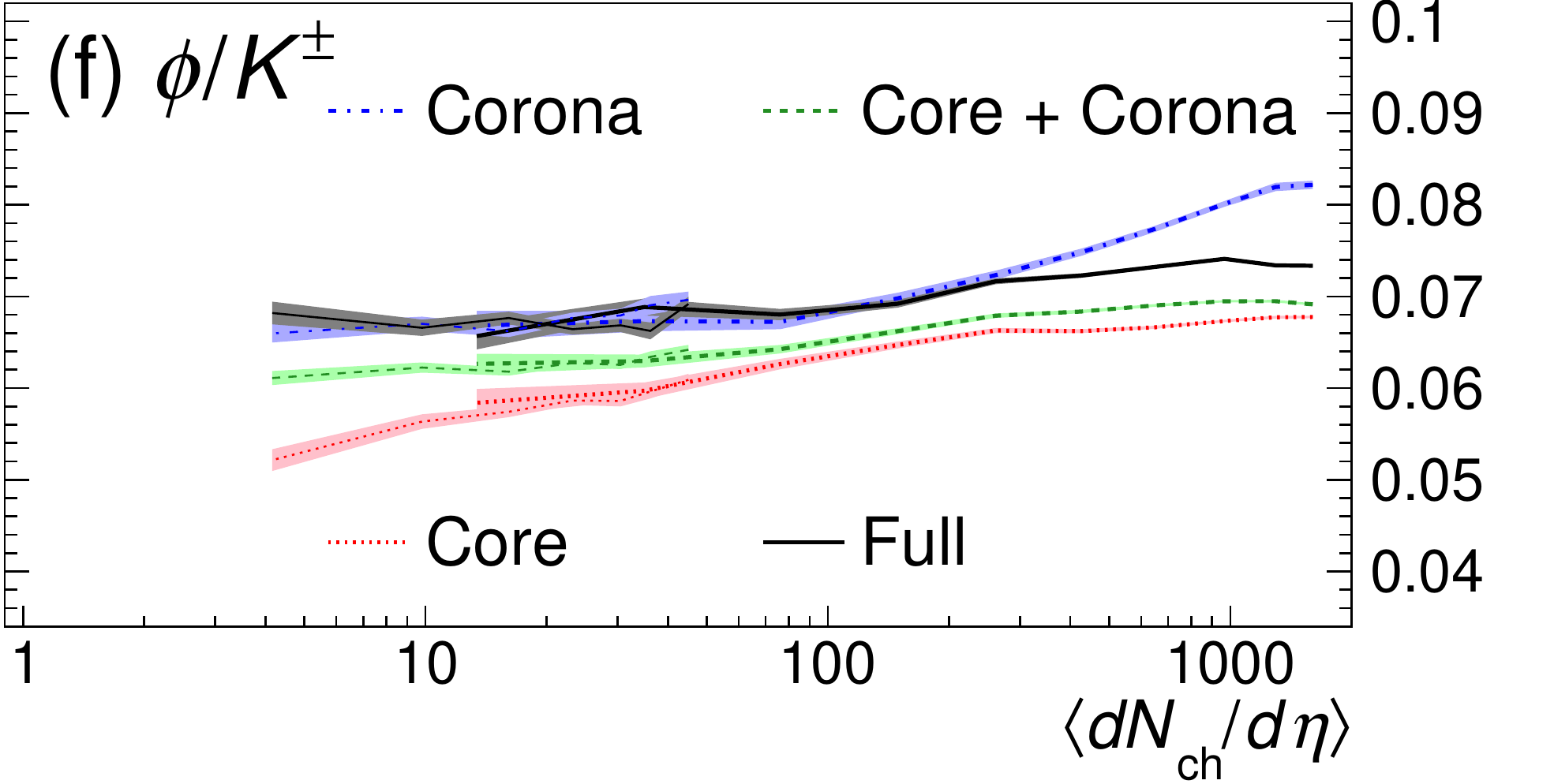}
\caption{Particle yield ratios in \ppb collisions at \rsnn[5.02~TeV] and \pb collisions at \rsnn~\cite{Knospe_EPOS_PbPb} (thin and thick lines, respectively). Results are shown for core only, corona only, summed core+corona (\textit{i.e.} UrQMD turned off), and full simulations (with core, corona, and UrQMD).}
\label{fig_coco}
\end{figure*}

While EPOS3 tends to overestimate the values of the ratios, it gives a good qualitative description of their system-size evolution. EPOS3 indicates no significant suppression of the \phk and \ssl ratios in \ppb and \ada collisions, consistent with ALICE~\cite{ALICE_Kstar_phi_pPb,ALICE_Kstar_phi_PbPb,ALICE_Kstar_phi_highpT_PbPb,ALICE_Sigmastar_Xistar_pPb} and STAR~\cite{STAR_Sigmastar_Lambdastar_AuAu} measurements. The ALICE data suggest a small multiplicity-dependent suppression of the \ksk ratio in \ppb collisions~\cite{ALICE_Kstar_phi_pPb} and a similar suppression may also be visible in a preliminary measurement of the \rhpi ratio in the same collision system~\cite{ALICE_rho_pPb}. ALICE measurements of the \ksk and \rhpi ratios in \pb collisions show a larger centrality-dependent suppression~\cite{ALICE_Kstar_phi_PbPb,ALICE_Kstar_phi_highpT_PbPb,ALICE_rho_PbPb}. These trends in \ppb and \pb collisions are qualitatively reproduced by EPOS3. The ALICE data indicate that the \lsl ratio does not change with multiplicity in \ppb collisions~\cite{ALICE_Lambdastar_pPb} but is suppressed in central \pb collisions~\cite{ALICE_Lambdastar_PbPb}; this trend is also qualitatively described by EPOS3. The ALICE measurement of the \xsx ratio is multiplicity-independent for \ppb collisions ~\cite{ALICE_Sigmastar_Xistar_pPb}, which are also consistent with a preliminary measurement in peripheral \pb collisions~\cite{Agrawal_SQM2017}. This behavior is also qualitatively reproduced by EPOS3. However, the preliminary ALICE measurement suggests a weak suppression of the \xsx ratio in (mid-)central \pb collisions (with values in the range 0.16--0.26); the magnitude of this suppression is not described by EPOS3. The EPOS3 model predicts that the \dlp ratio should not depend on multiplicity in \ppb collisions, which is a reasonable expectation in light of STAR's measurement~\cite{STAR_resonances_dAu} of this ratio in $d$-Au collisions. In summary, the multiplicity evolution of these various ratios are qualitatively well described by EPOS3, with the possible exception of the \xsx ratio in large collision systems. Furthermore, we observe smooth evolution of the particle yield ratios from the lowest multiplicity \ppb collisions to central \pb collisions, with little or no difference between \ppb and \pb collisions at similar charged-particle multiplicities.

The role played by the resonance lifetime should be noted. The two resonances with the clearest suppression, \rh and \ks, are both short-lived. In contrast, the \ph has a long lifetime and is not suppressed. \ls and \xs have intermediate lifetimes within the range considered, and neither is suppressed in \ppb collisions. In \pb collisions, \ls is suppressed and there is weaker suppression of \xs, with a lifetime approximately half that of the \ph and twice that of the \ls. However, the \ssx and \dl are short-lived, but are not suppressed (indeed, they are enhanced from low to high multiplicity beyond the statistical uncertainties of the EPOS3 calculations). Taken together, these results indicate that while the lifetime is an important factor in determining whether a resonance yield is suppressed, it is not the only factor. One must also account for (1) the various scattering cross-sections of the decay products; (2) the different $Q$ values of the decays; (3) the complicated interplay among re-scattering, regeneration, and feed-down; and (4) the interplay between the core and corona parts of the collision.

\begin{figure*}[!ht]
\centering
\includegraphics[angle=0,scale=0.29]{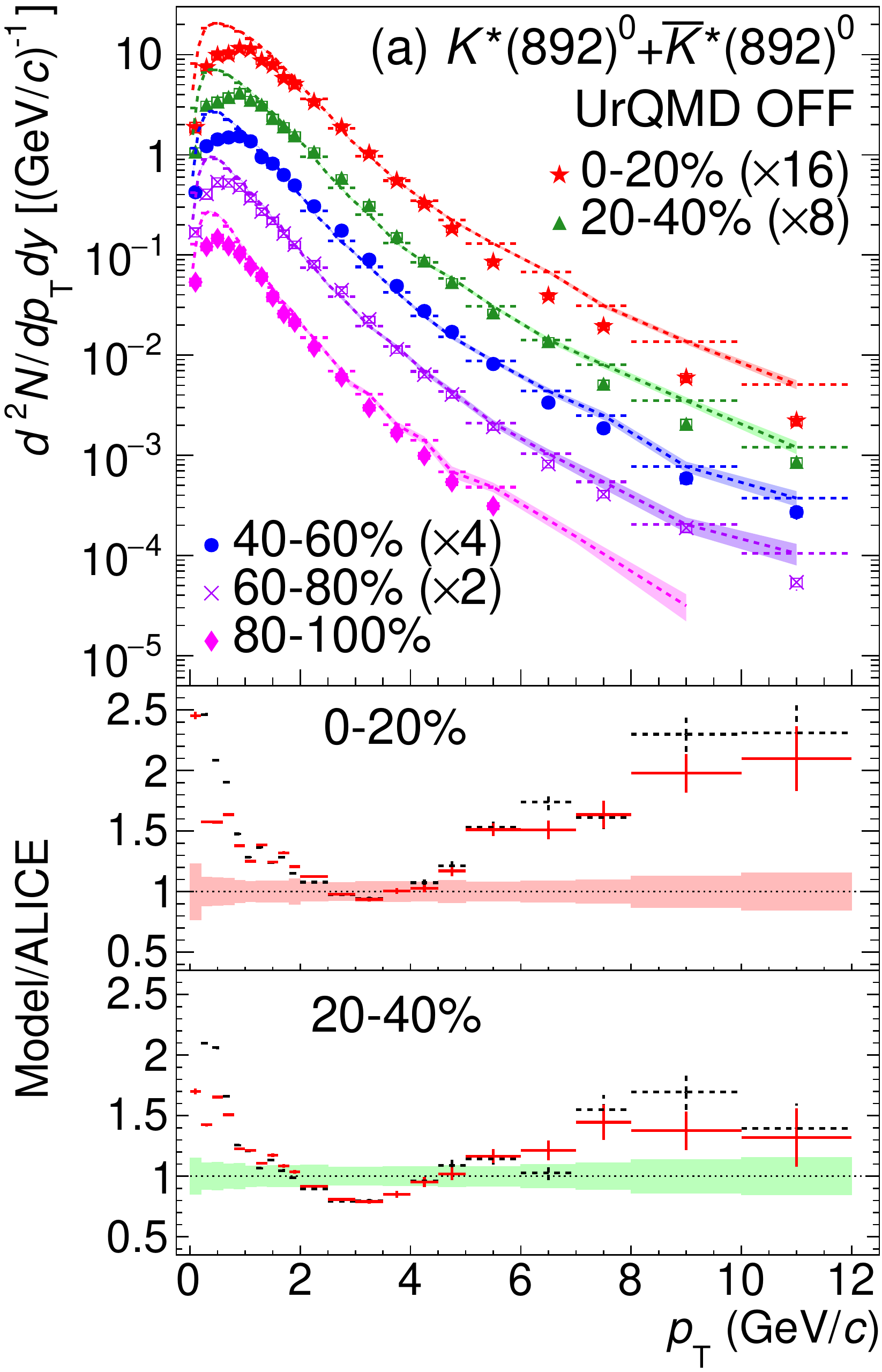}
\includegraphics[angle=0,scale=0.29]{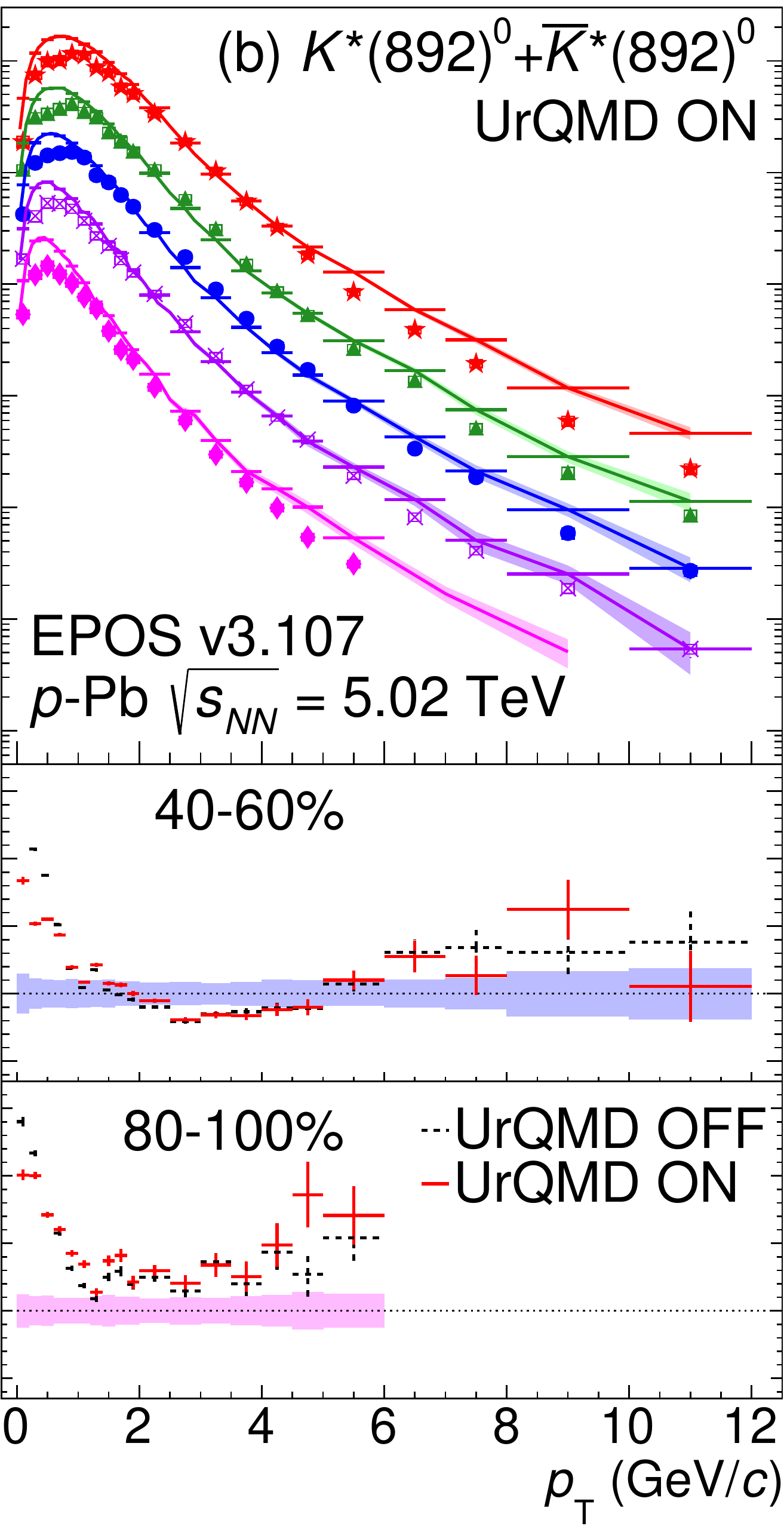}
\includegraphics[angle=0,scale=0.29]{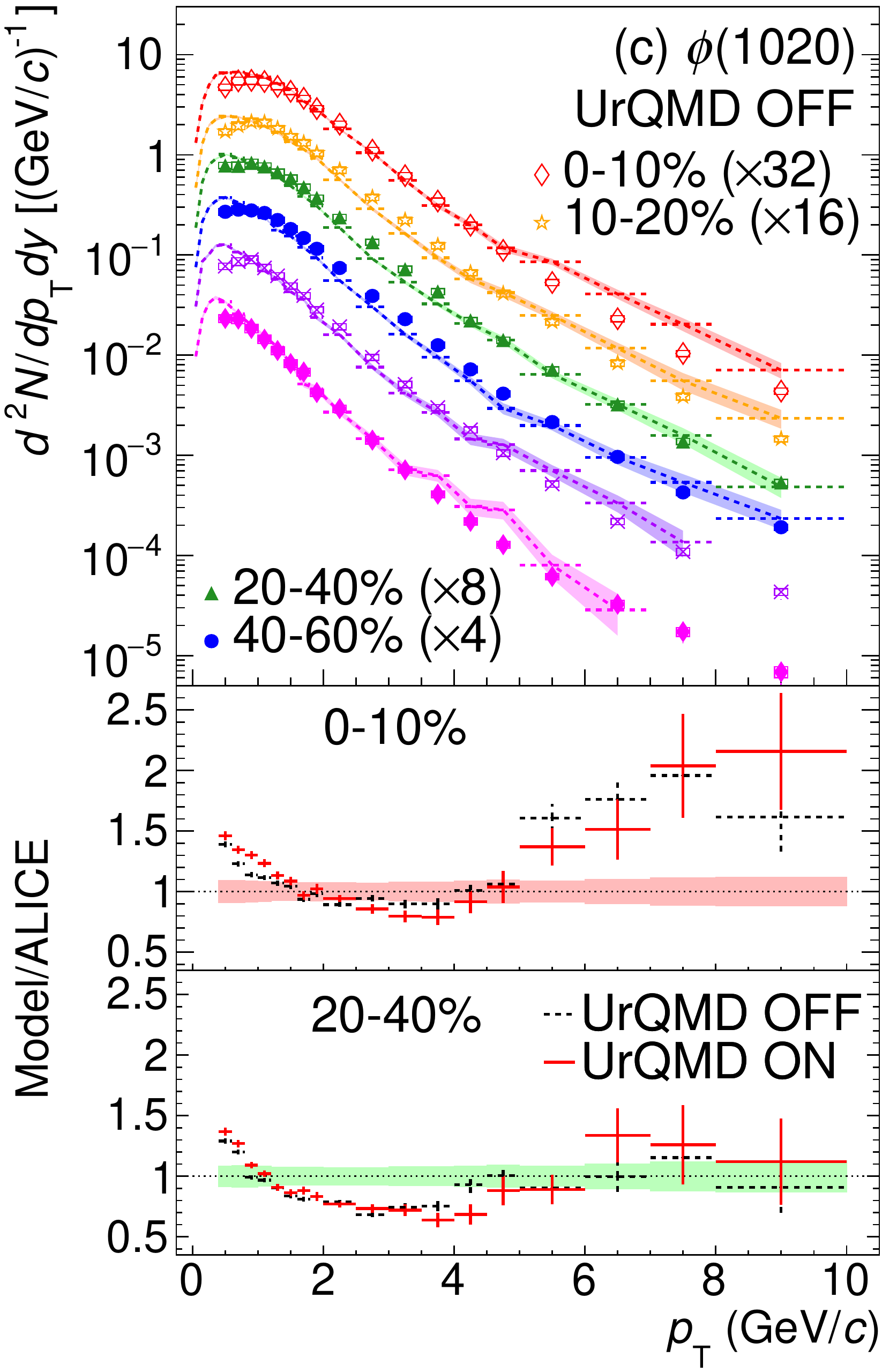}
\includegraphics[angle=0,scale=0.29]{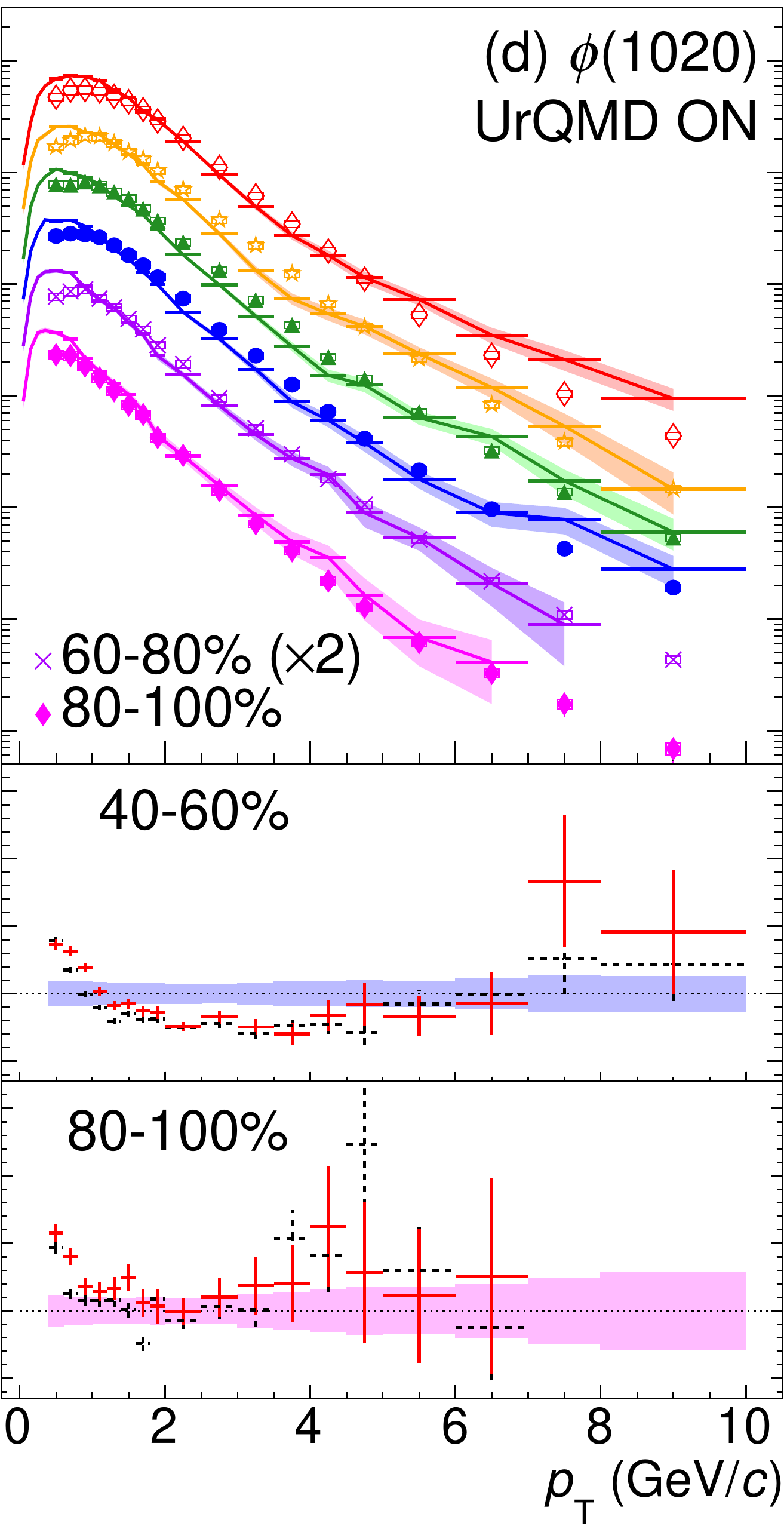}\\
\includegraphics[angle=0,scale=0.29]{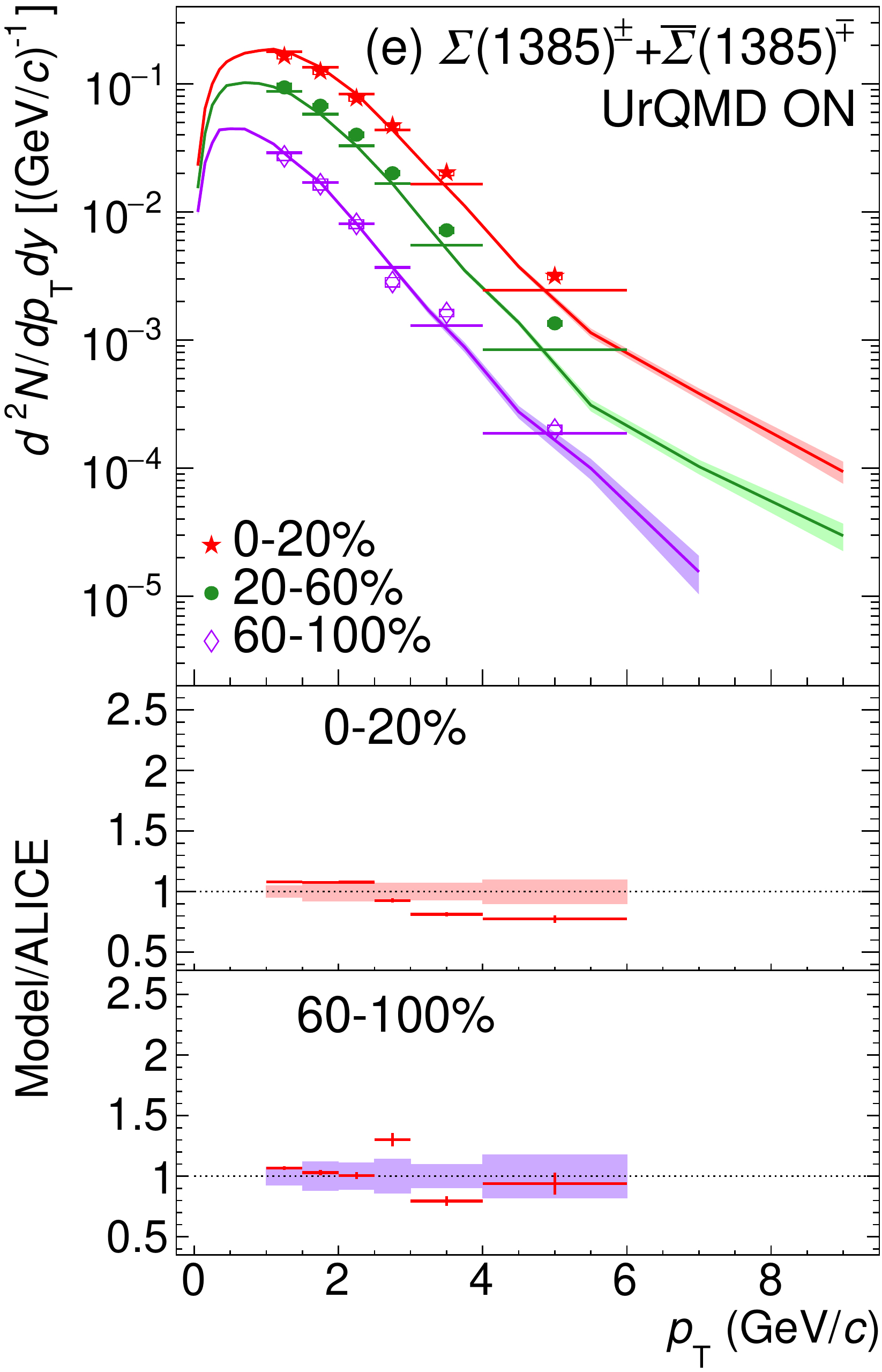}
\includegraphics[angle=0,scale=0.29]{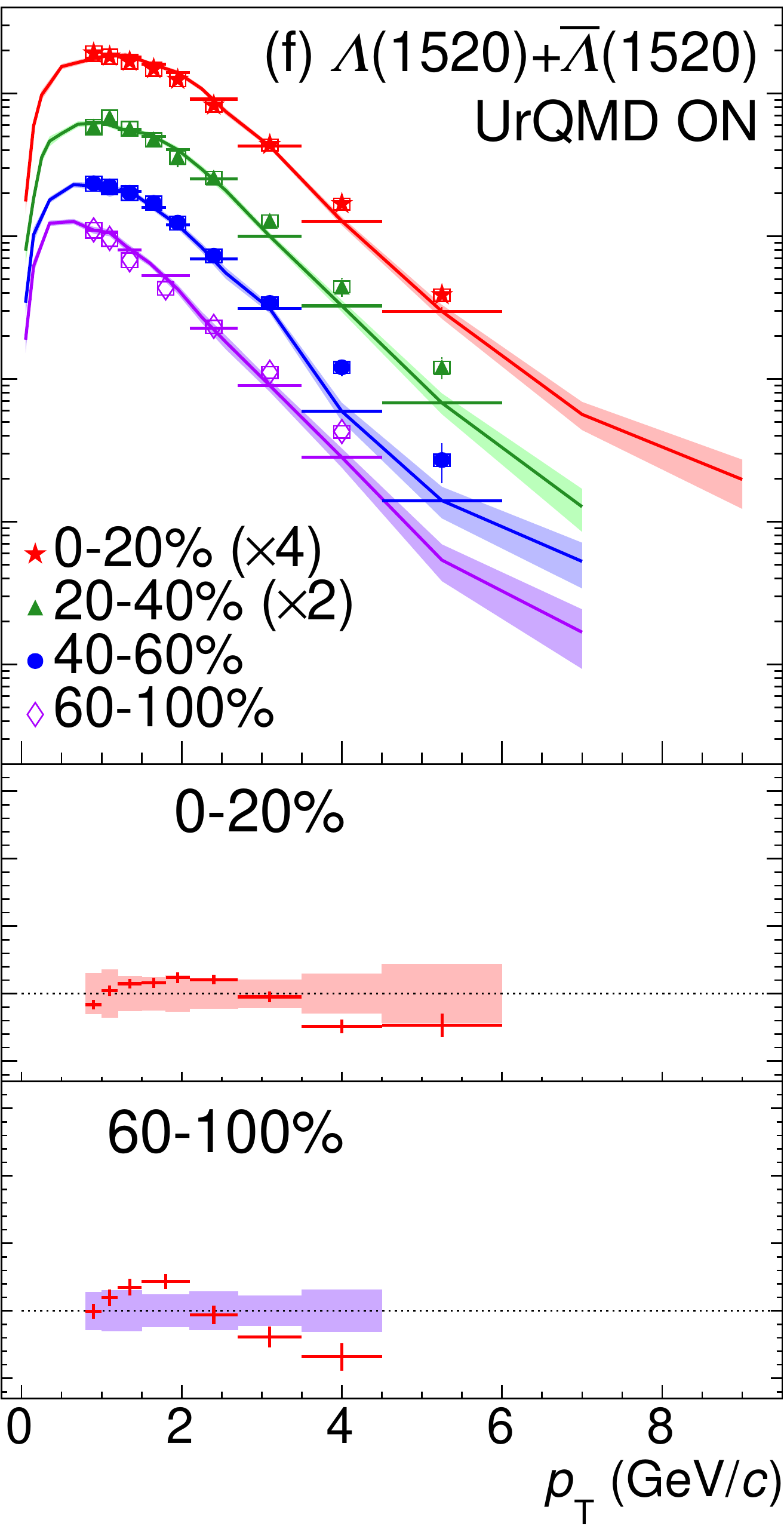}
\includegraphics[angle=0,scale=0.29]{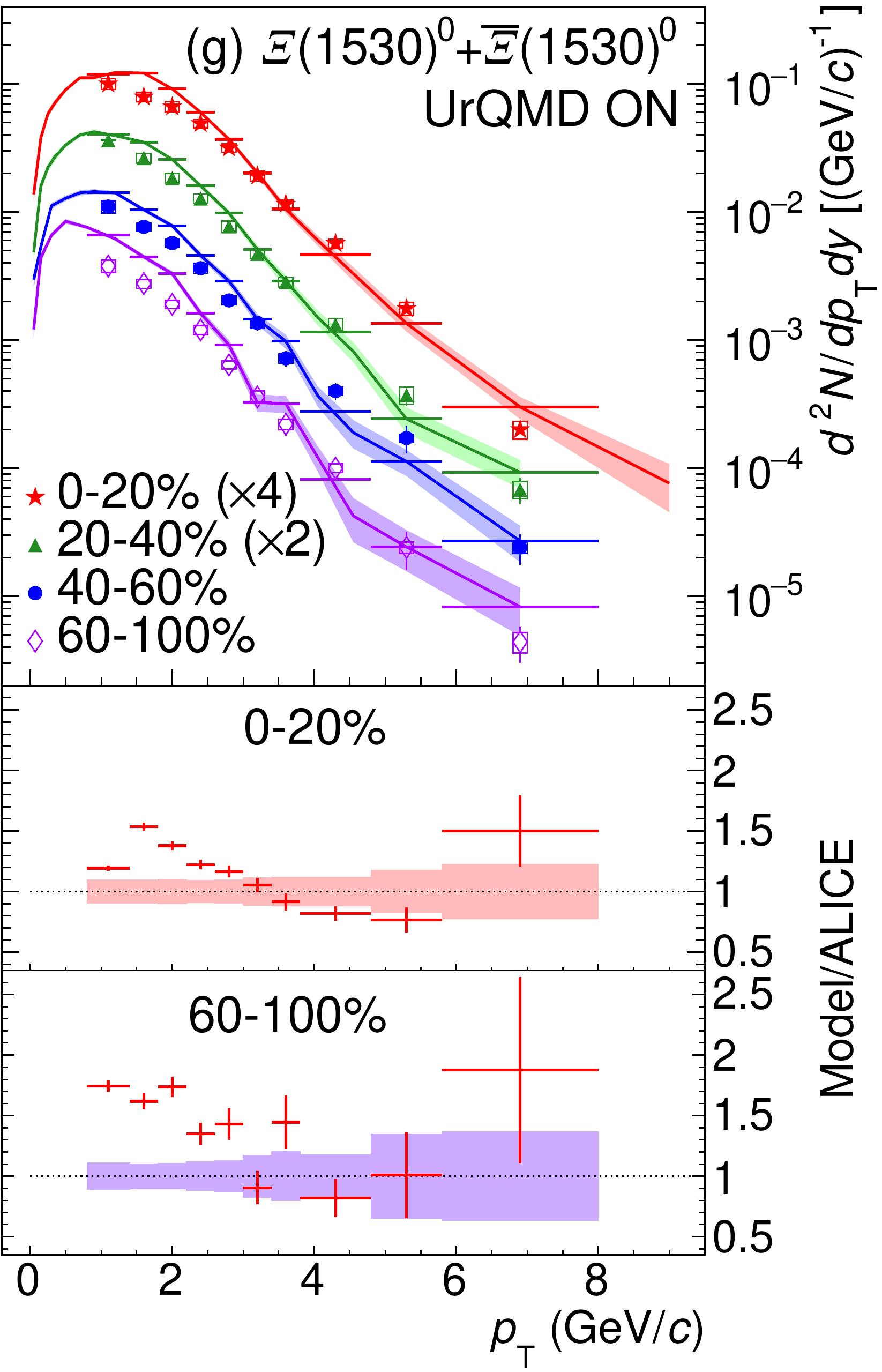}
\caption{Comparison of EPOS3 results and ALICE measurements for resonances in \ppb collisions at \rsnnppb for various multiplicity intervals. The curves are the EPOS3 \pT distributions with fine bins and the shaded bands are the statistical uncertainties of the EPOS3 results. The horizontal lines are the EPOS3 results in the same \pT bins as the ALICE measurements. \textbf{Lower subpanels:} the ratios of the EPOS3 results to the ALICE measurements as functions of \pT for the different centrality intervals. The shaded bands around unity represent the uncertainties of the measured data. \textbf{(a-b):} $\ks+\aks$~\cite{ALICE_Kstar_phi_pPb} without (a) and with (b) UrQMD. The ALICE data are the same in both panels. \textbf{(c-d):} \ph~\cite{ALICE_Kstar_phi_pPb} without (c) and with (d) UrQMD. The ALICE data are the same in both panels. \textbf{(e):} $\ssx+\assx$~\cite{ALICE_Sigmastar_Xistar_pPb}, \textbf{(f):} $\ls+\als$~\cite{ALICE_Lambdastar_pPb}, \textbf{(g):} $\xs+\axs$~\cite{ALICE_Sigmastar_Xistar_pPb}.
}
\label{fig_spectra}
\end{figure*}

\begin{figure*}[!ht]
\centering
\includegraphics[angle=0,scale=0.29]{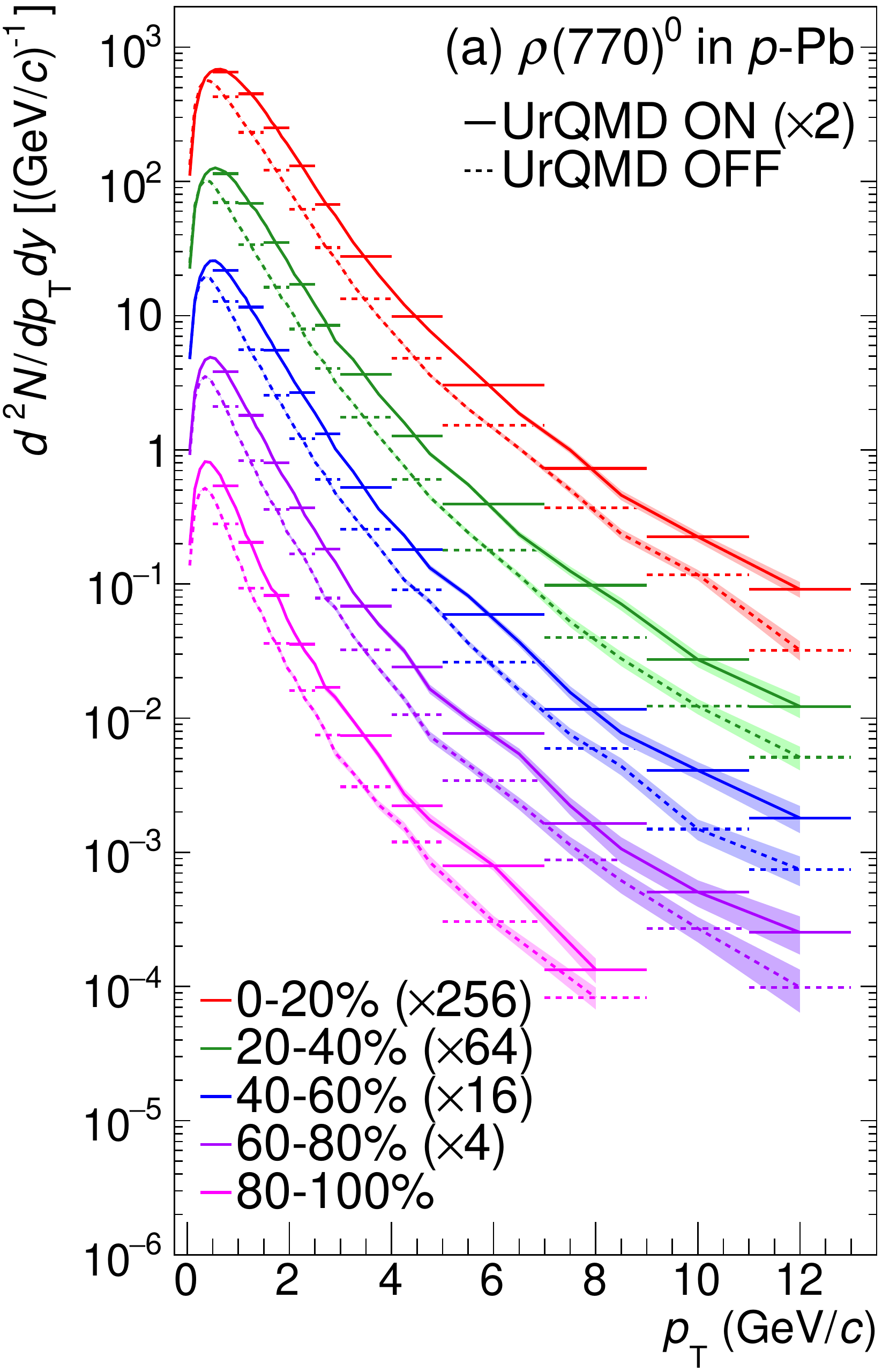}
\includegraphics[angle=0,scale=0.29]{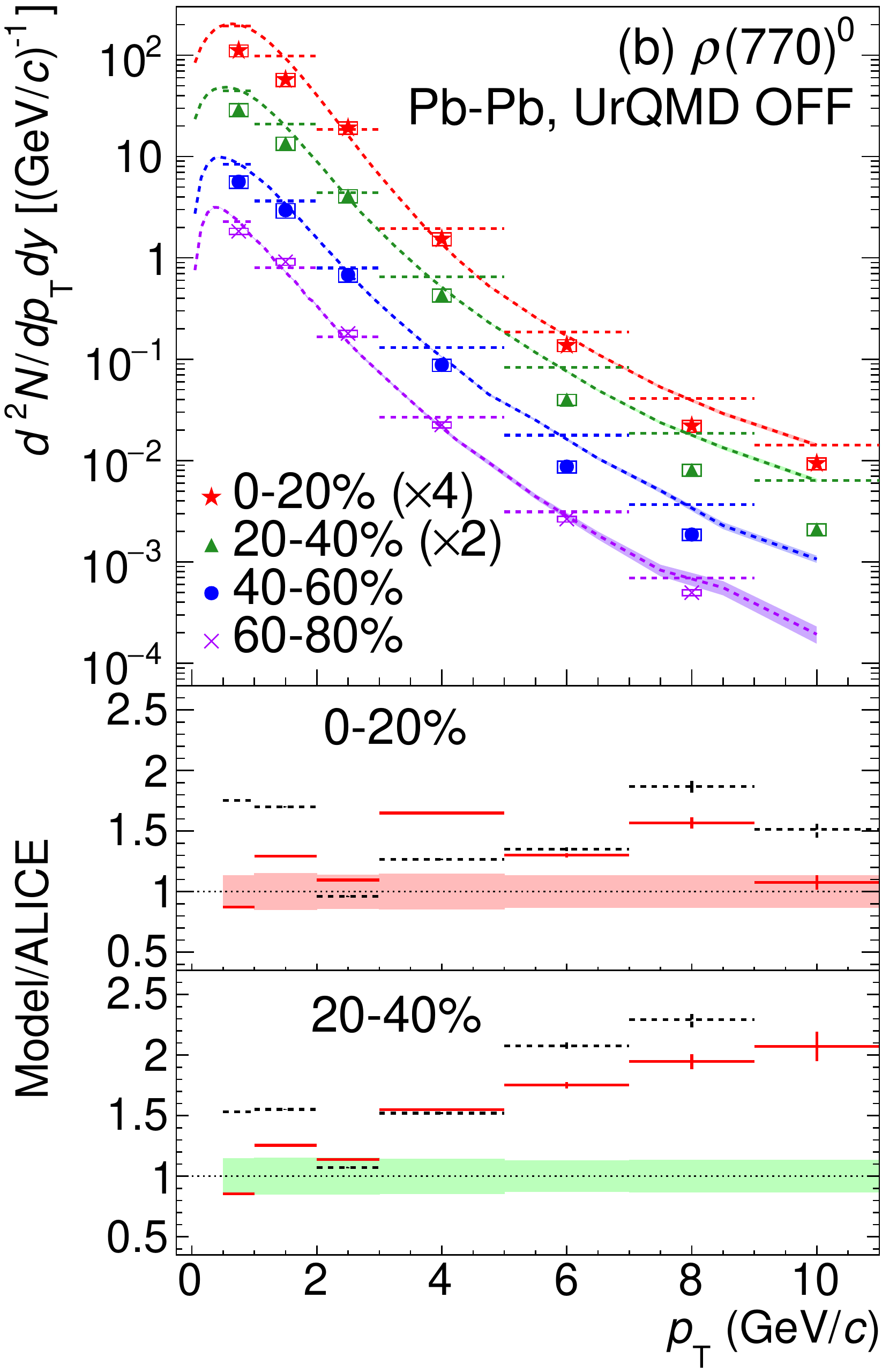}
\includegraphics[angle=0,scale=0.29]{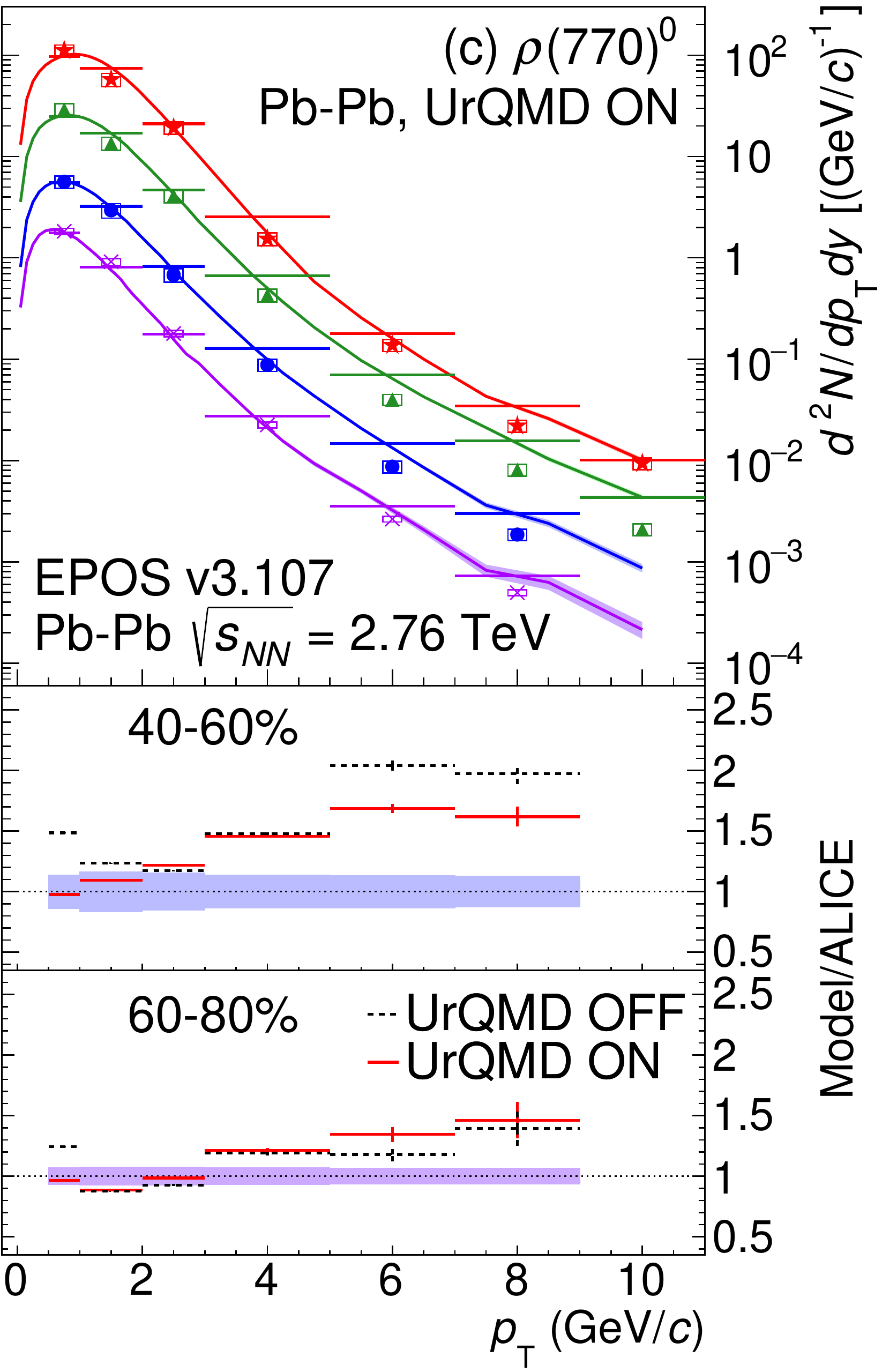}
\caption{\textbf{(a):} EPOS3 \pT spectra for \rh mesons in \ppb at \rsnnppb in various multiplicity intervals with UrQMD off (dashed lines) and UrQMD on (solid lines, scaled by an extra factor of 2). The curves are the EPOS3 \pT distributions with fine bins and the shaded bands are the statistical uncertainties of the EPOS3 results. The horizontal lines are the EPOS3 results in wider \pT bins. \textbf{(b-c):} Comparison of EPOS3 results~\cite{Knospe_EPOS_PbPb} and ALICE measurements~\cite{ALICE_rho_PbPb} for the \rh meson in \pb collisions at \rsnn[2.76]~TeV in various centrality intervals. \textbf{Upper panels:} transverse momentum distributions of \rh from EPOS3 with UrQMD OFF (left) and UrQMD ON (right). These results are compared to measurements from the ALICE Experiment (same data on left and right). The curves are the EPOS3 \pT distributions with fine bins and the shaded bands are the statistical uncertainties of the EPOS3 results.  The horizontal lines are the EPOS3 results in the same \pT bins as the ALICE measurements. \textbf{Lower panels:} the ratio of the EPOS3 results to the ALICE measurements as functions of \pT for the different centrality intervals. The shaded bands around unity represent the uncertainties of the measured data.}
\label{fig_rho_spectra_PbPb}
\end{figure*}

\begin{figure*}[!ht]
\centering
\includegraphics[angle=0,scale=0.75]{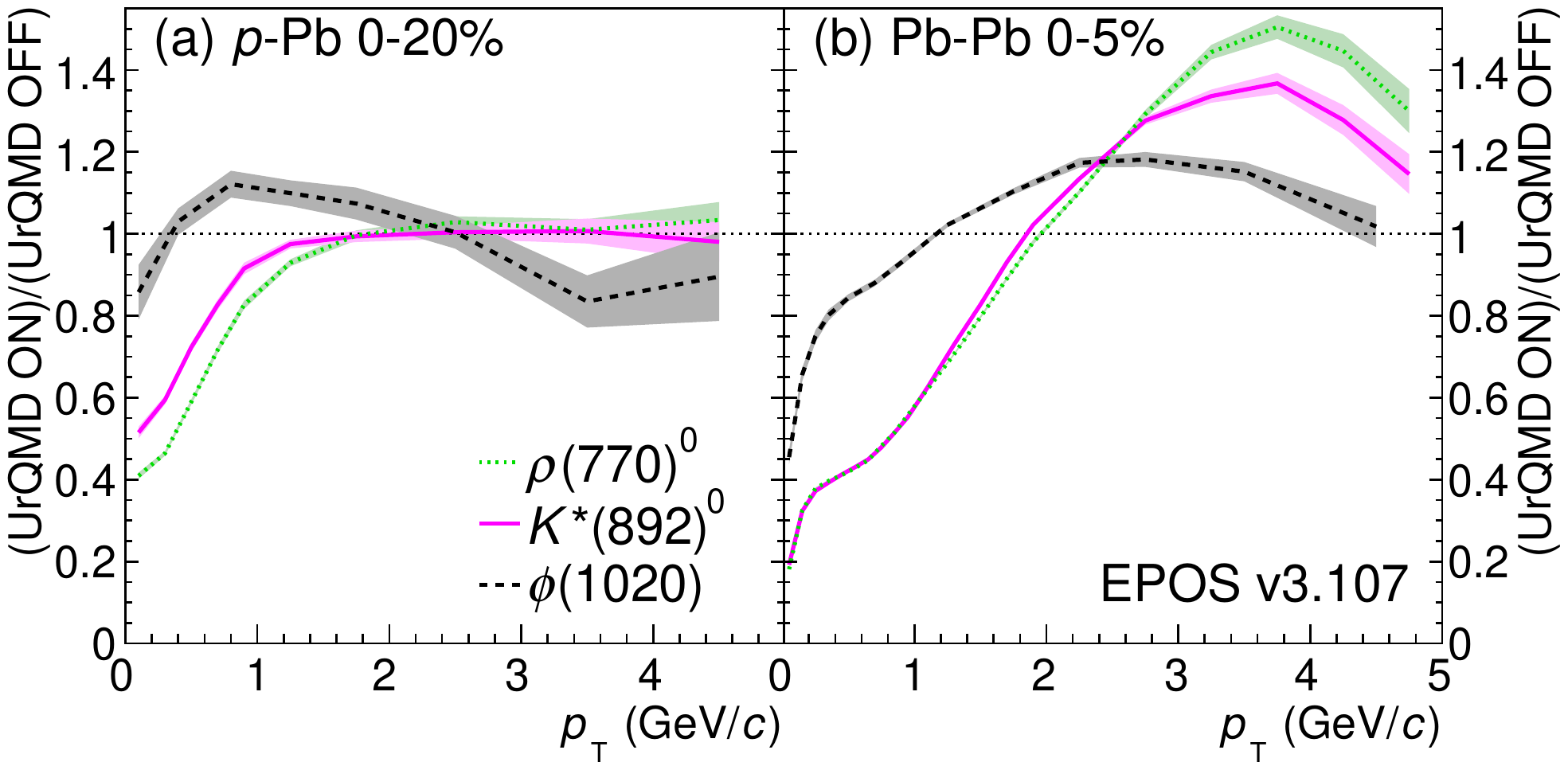}
\caption{\textbf{(a):} The effect of the UrQMD phase on the \pT spectra of the \rh, \ks, and \ph mesons is illustrated by this \pT-dependent ratio: the \pT spectra from EPOS3 with UrQMD ON divided by the spectra with UrQMD OFF. Results are shown for (a) high-multiplicity \ppb collisions at \rsnnppb and (b) central \pb collisions at \rsnn~\cite{Knospe_EPOS_PbPb}. The shaded bands represent the statistical uncertainties of the ratio.}
\label{fig_UrQMD_ratio_compare}
\end{figure*}

Figure~\ref{fig_coco} shows the system-size evolution of various particle ratios given by EPOS3 for \ppb and \pb collisions. The figure shows separately the ratios for particles produced in the core and in the corona, as well as for the combined core+corona system. The core+corona ratio is then further modified by interactions in the hadronic phase to give the final ratios. The core+corona ratio can be viewed as an interpolation between the extreme cases: the corona-only and core-only ratios. The system is mostly corona in small collisions and mostly core in large collisions, so the core+corona ratio can evolve with multiplicity even if the corona-only or core-only ratios are multiplicity-independent (\textit{e.g.} the K/$\pi^{\pm}$ ratio).

The decreases in the \rhpi and \ksk ratios are partly explained by the core-corona transition, with further suppression coming from UrQMD. Turning on UrQMD causes a small increase in the \phk ratio for all multiplicities. The suppression in \ppb collisions of the yields of short-lived resonances is therefore partly explained by the existence of a hadronic phase with a non-zero lifetime (about 2 fm/$c$ in high-multiplicity \ppb collisions). The question of whether an extended partonic phase is the precursor for the extended hadronic phase is not answered here. It is also interesting to note the effect of the hadronic phase on the $p/\pi^{\pm}$ and $\Lambda/\pi^{\pm}$ ratios: interactions in the hadronic phase cause a significant suppression of the proton and $\Lambda$ yields, especially in central \ada collisions, due to particle-antiparticle annihilation and absorption effects.

\section{Resonance Transverse Momentum Distributions}

Figures~\ref{fig_spectra} and \ref{fig_rho_spectra_PbPb} show the \pT distributions of various resonances, as calculated using EPOS3 and measured by ALICE~\cite{ALICE_Kstar_phi_pPb,ALICE_Sigmastar_Xistar_pPb,ALICE_Lambdastar_pPb,ALICE_rho_PbPb}. The \pT spectra of \ks and \ph in \ppb collisions at \rsnn[5.02]~TeV are shown in Fig.~\ref{fig_spectra}. In general, the EPOS3 spectra are softer (steeper negative slopes) than the measured spectra at low \pT and harder (flatter slopes) at high \pT, with the best qualitative description of the shapes in the approximate range $2\lesssim\pT\lesssim4$~\gvc. The agreement between the EPOS3 calculations and the measured spectra improves for lower multiplicity \ppb collisions. EPOS3 calculations are shown with UrQMD (``UrQMD ON") and without it (``UrQMD OFF"). The effect of turning UrQMD on is greatest for $\pT\lesssim1$~\gvc, resulting in a notable improvement in the description of the \ks meson \pT spectra at low \pT. This behavior is consistent with the expectation that re-scattering should be most important at low momenta. The effects of UrQMD on \ph \pT spectra are fairly small. Similar behavior was observed for these resonances in \pb collisions.

Figure~\ref{fig_spectra} also shows comparisons of EPOS3 calculations (with UrQMD on) and ALICE measurements~\cite{ALICE_Sigmastar_Xistar_pPb,ALICE_Lambdastar_pPb} of the \pT spectra of three baryonic resonances in \ppb collisions at \rsnnppb: \ssx, \xs, and \ls. The EPOS3 spectra tend to be softer than the measured ALICE spectra. The EPOS3 spectra for \ssx and \ls provide fair descriptions of the ALICE data, while EPOS3 overestimates the \xs yields at low \pT.

Figure~\ref{fig_rho_spectra_PbPb} shows the \pT spectra given by EPOS3 in \ppb collisions at \rsnnppb for future comparison with experimental results. Note that turning on UrQMD leads to a large suppression of the \rh yield at very low \pT ($\pT<0.5$~\gvc) Figure~\ref{fig_rho_spectra_PbPb} also shows a comparison of the EPOS3 and ALICE \pT spectra for \rh mesons in \pb collisions at \rsnn[2.76]~TeV~\cite{ALICE_rho_PbPb}. As for our previous studies of the \ks and \ph in \pb collisions~\cite{Knospe_EPOS_PbPb}, turning UrQMD on improves the agreement between the EPOS3 calculation and the ALICE data at low \pT, while the EPOS3 description of the spectra improves for more peripheral collisions.

Figure~\ref{fig_UrQMD_ratio_compare} shows the effect of UrQMD on the \pT spectra of the \rh, \ks, and \ph resonances in both \ppb and \ppb collisions. The effect is quantified by a ratio: the \pT spectra obtained from the event samples produced with UrQMD are divided by the corresponding \pT spectra produced without UrQMD. For both collision systems, a depletion is observed at low \pT, below \mbox{1-2}~\gvc. This depletion is due to scattering effects, but may be partly filled in due to regeneration. The depletion is much more pronounced for \pb collisions than for \ppb and larger for the shorter lived resonances \rh and \ks than for \ph. There is also an enhancement of these resonances for intermediate \pT in \pb collisions.

\begin{figure*}[!ht]
\centering
\includegraphics[angle=0,scale=0.42]{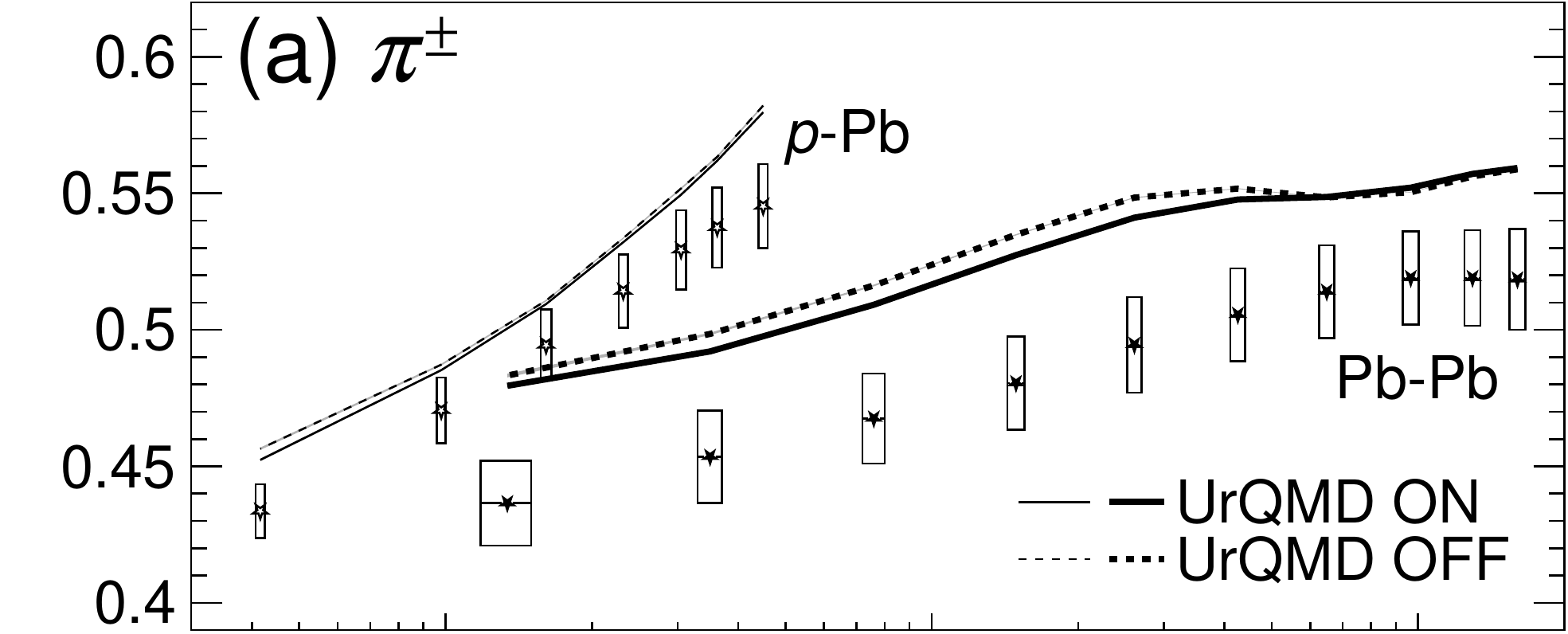}
\includegraphics[angle=0,scale=0.42]{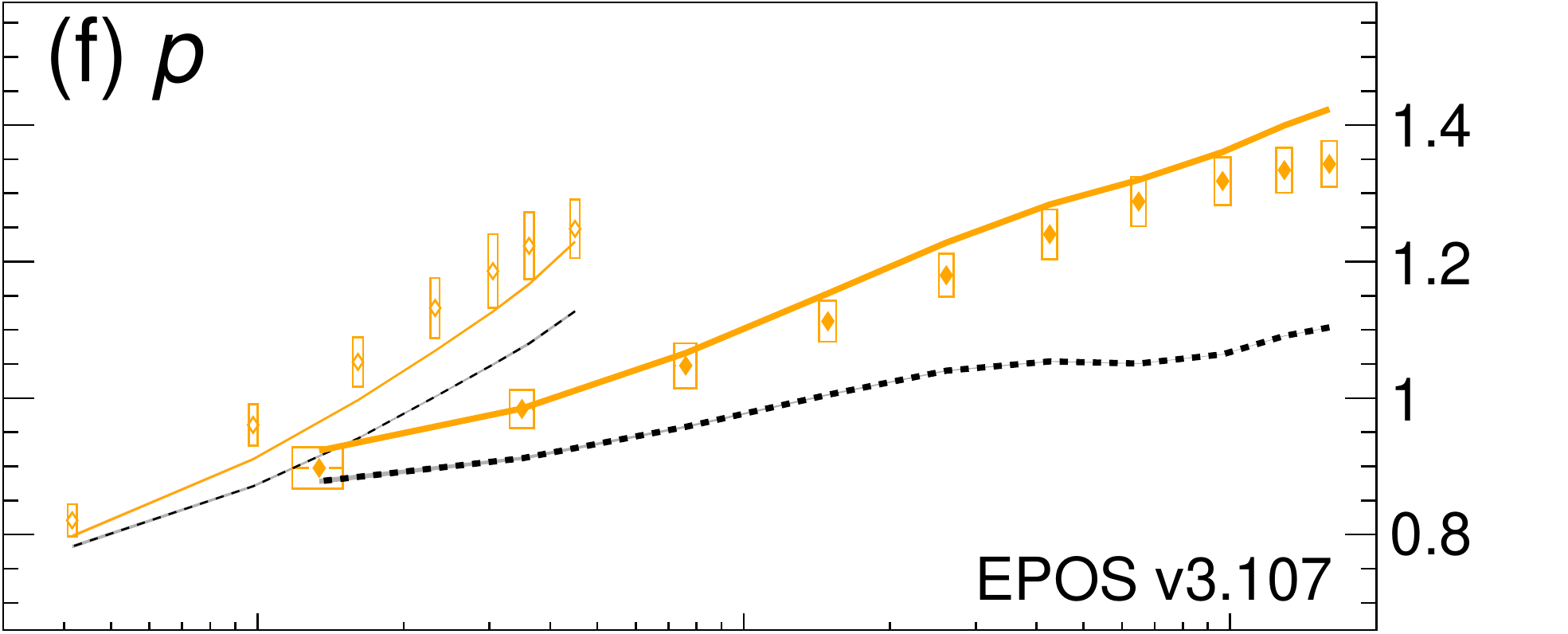}
\includegraphics[angle=0,scale=0.42]{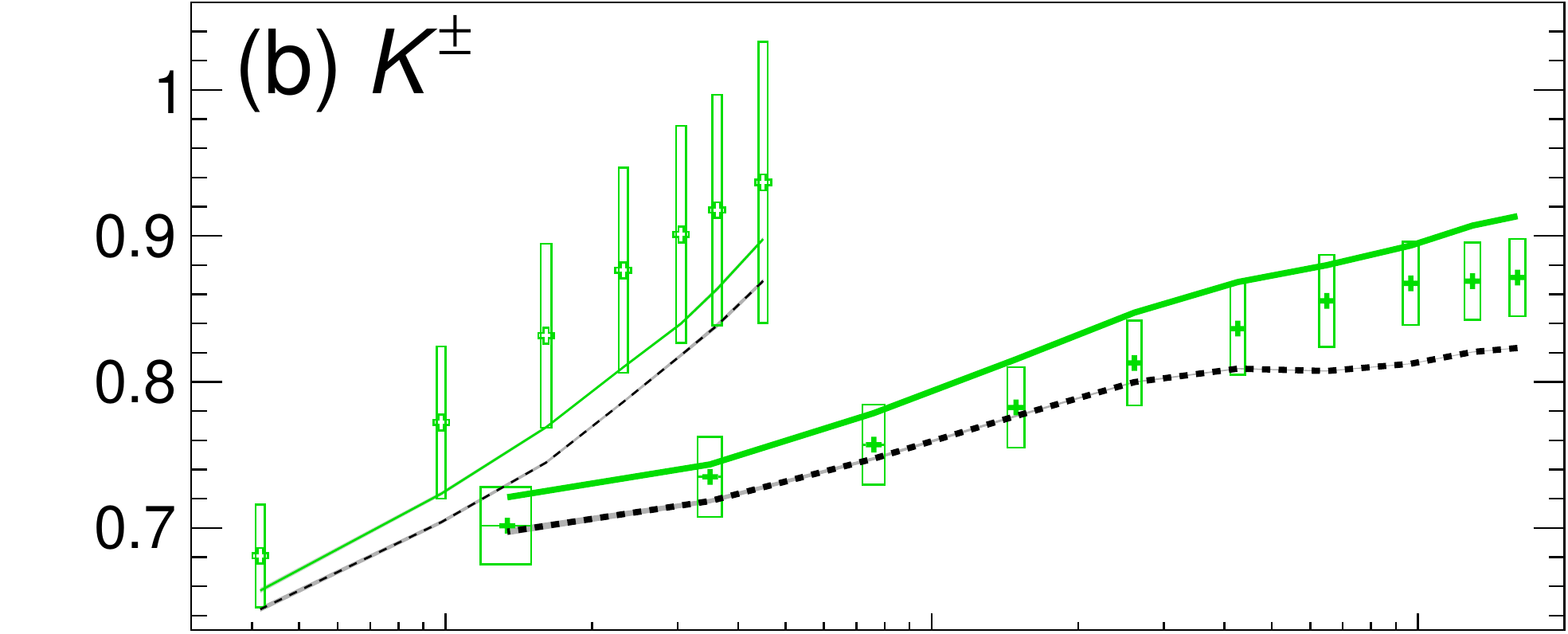}
\includegraphics[angle=0,scale=0.42]{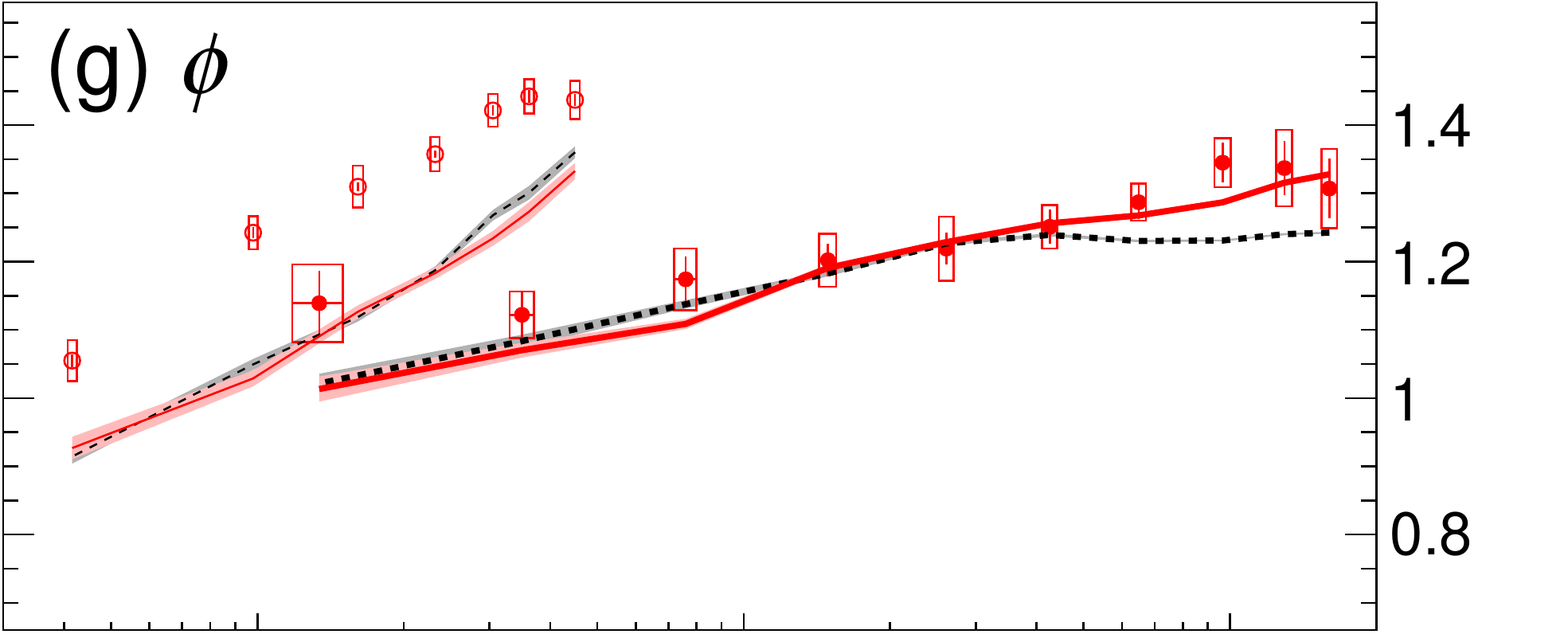}
\includegraphics[angle=0,scale=0.42]{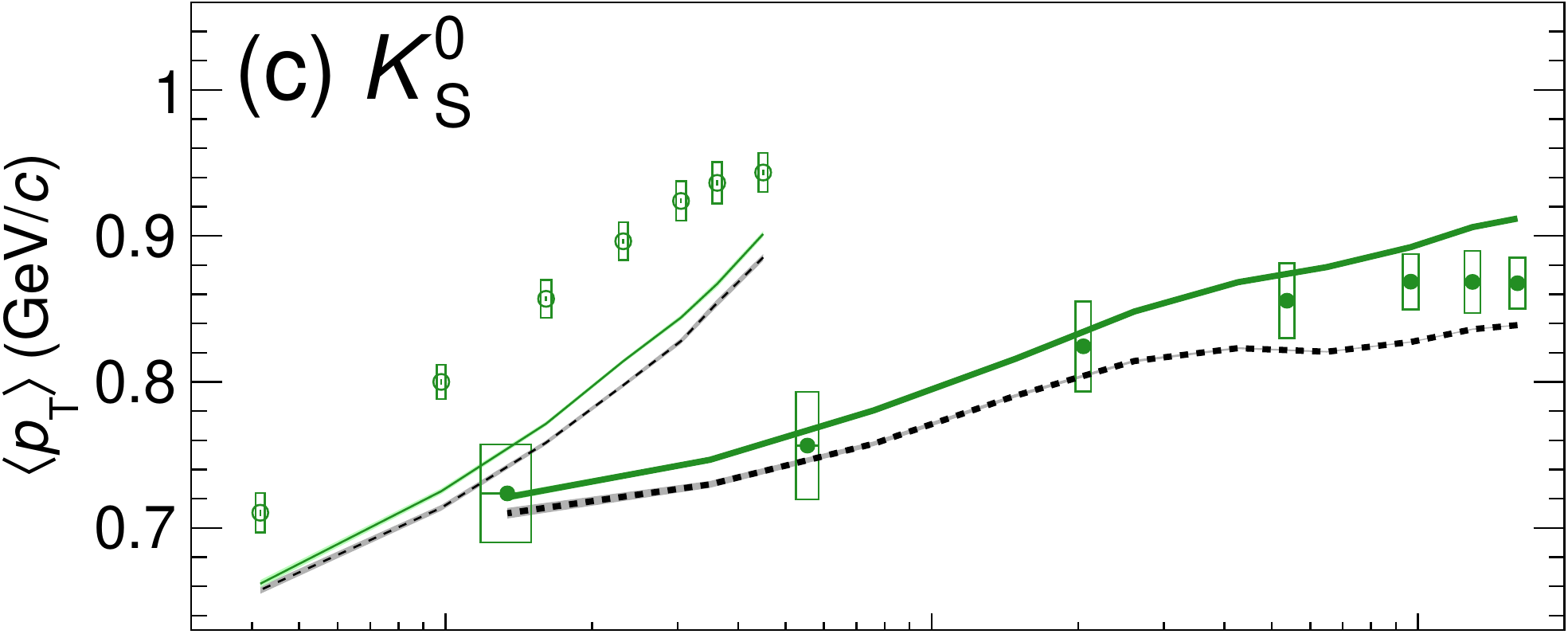}
\includegraphics[angle=0,scale=0.42]{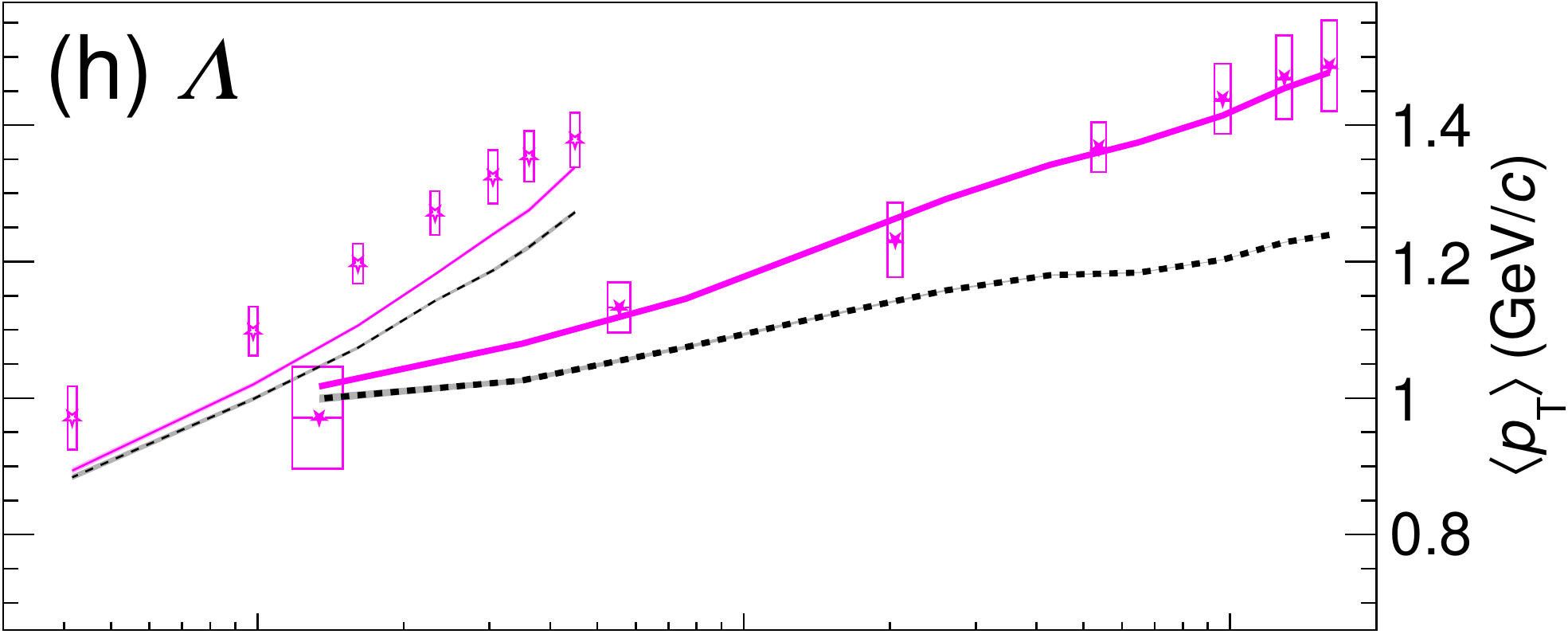}
\includegraphics[angle=0,scale=0.42]{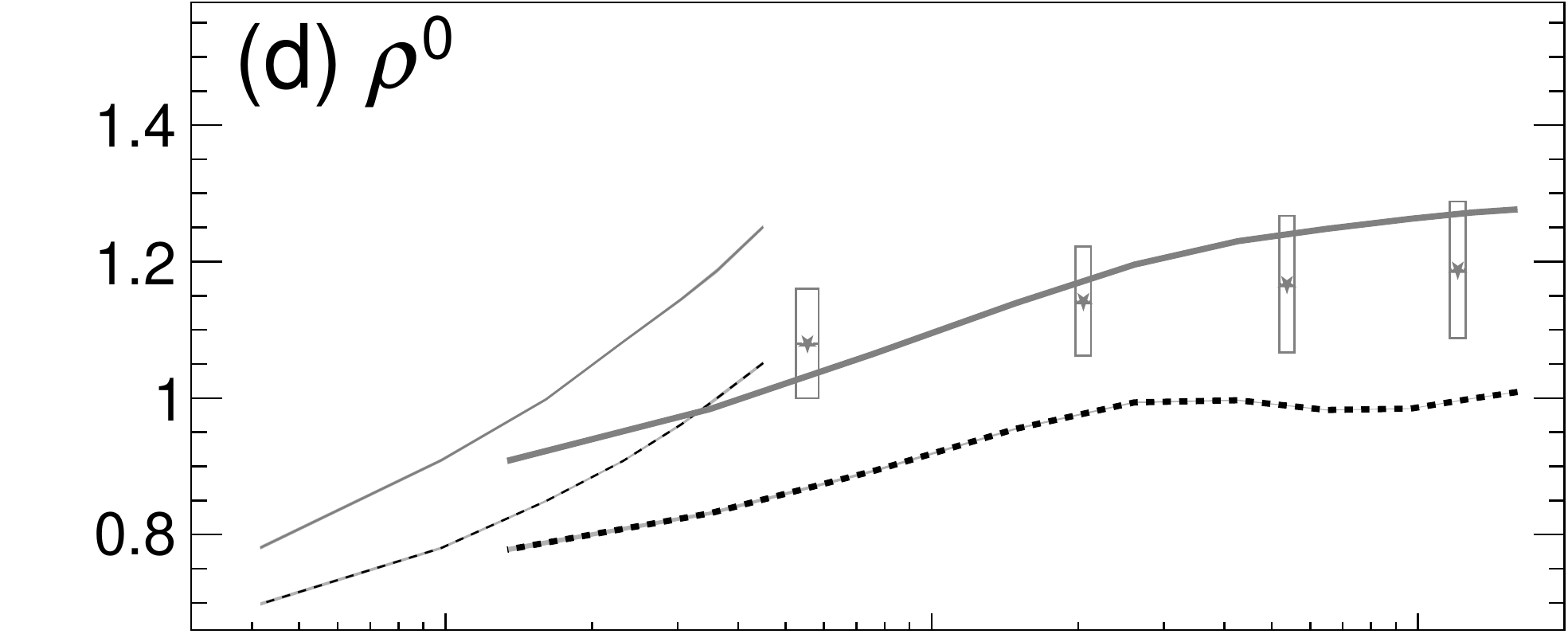}
\includegraphics[angle=0,scale=0.42]{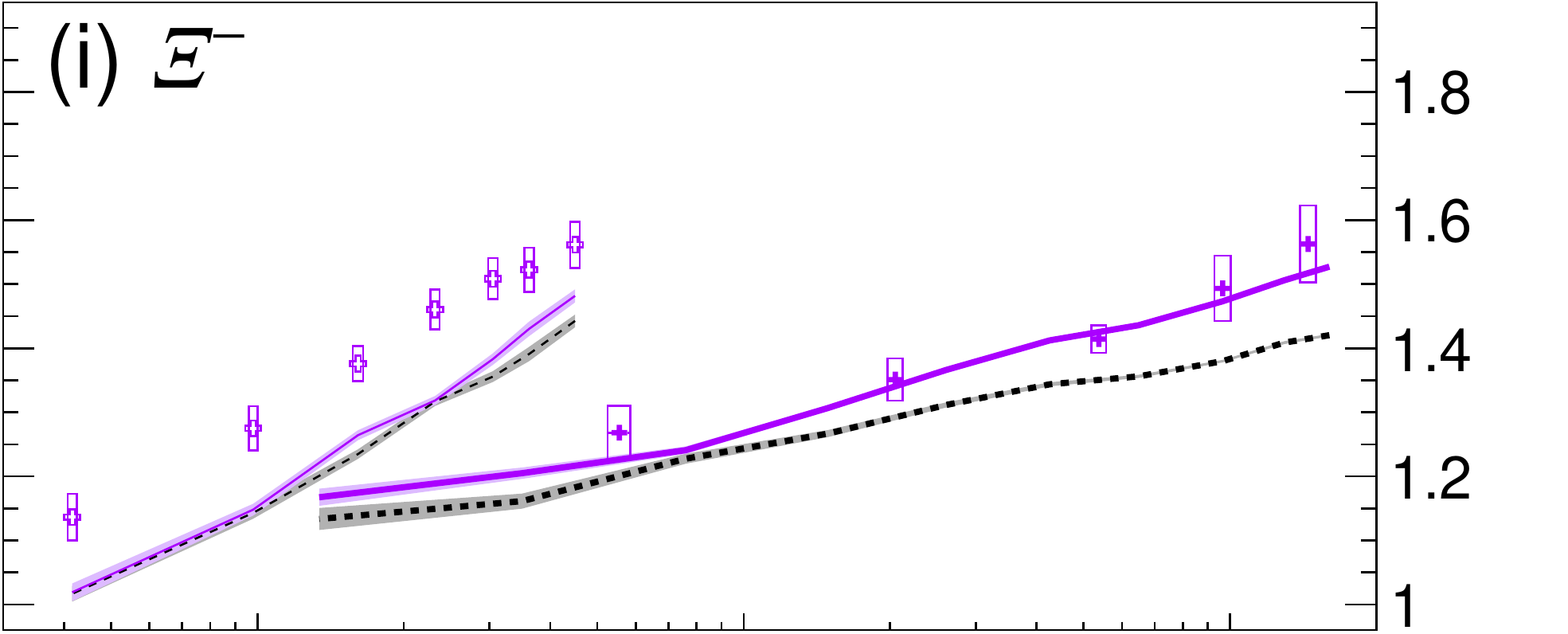}
\includegraphics[angle=0,scale=0.42]{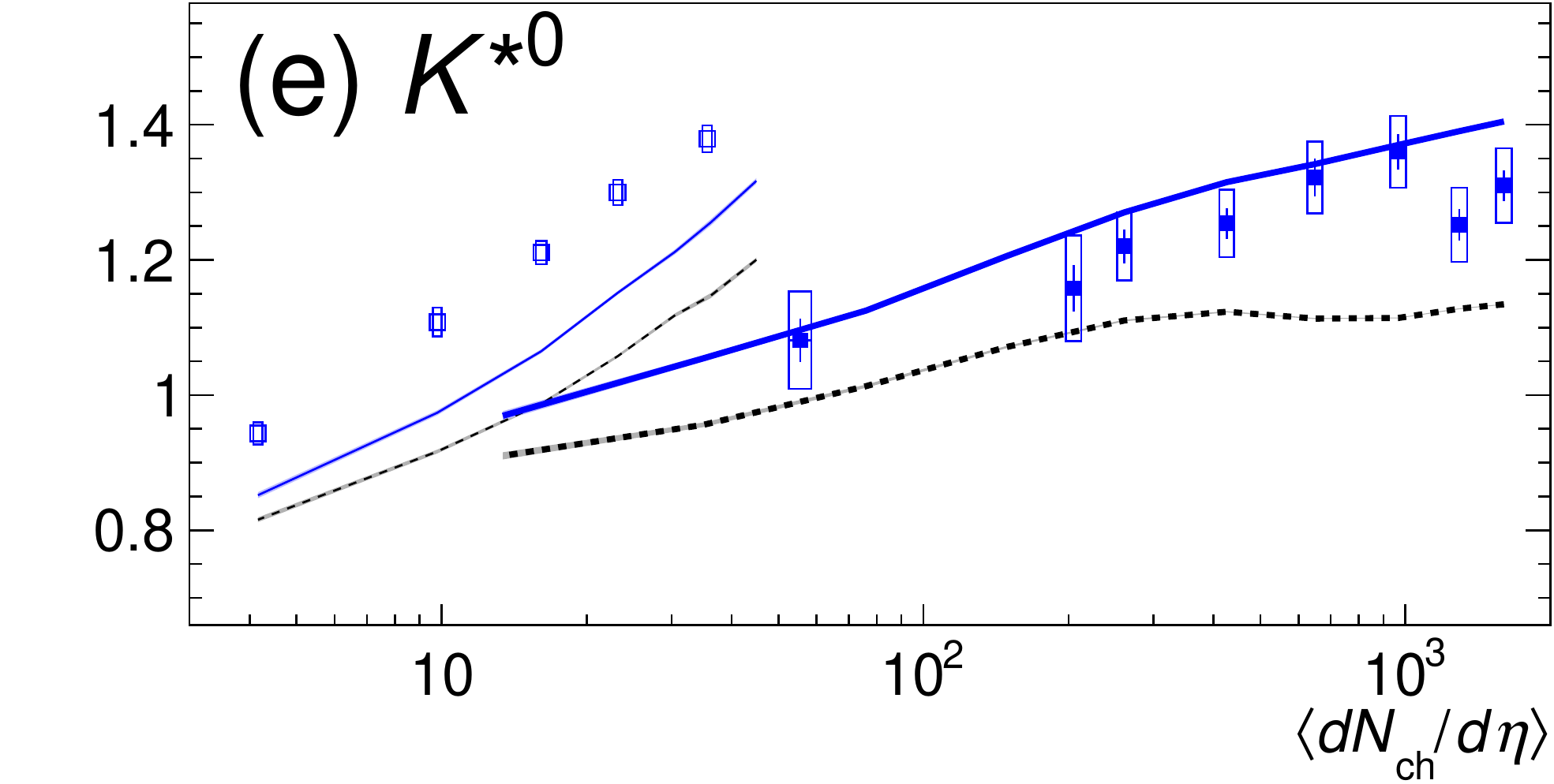}
\includegraphics[angle=0,scale=0.42]{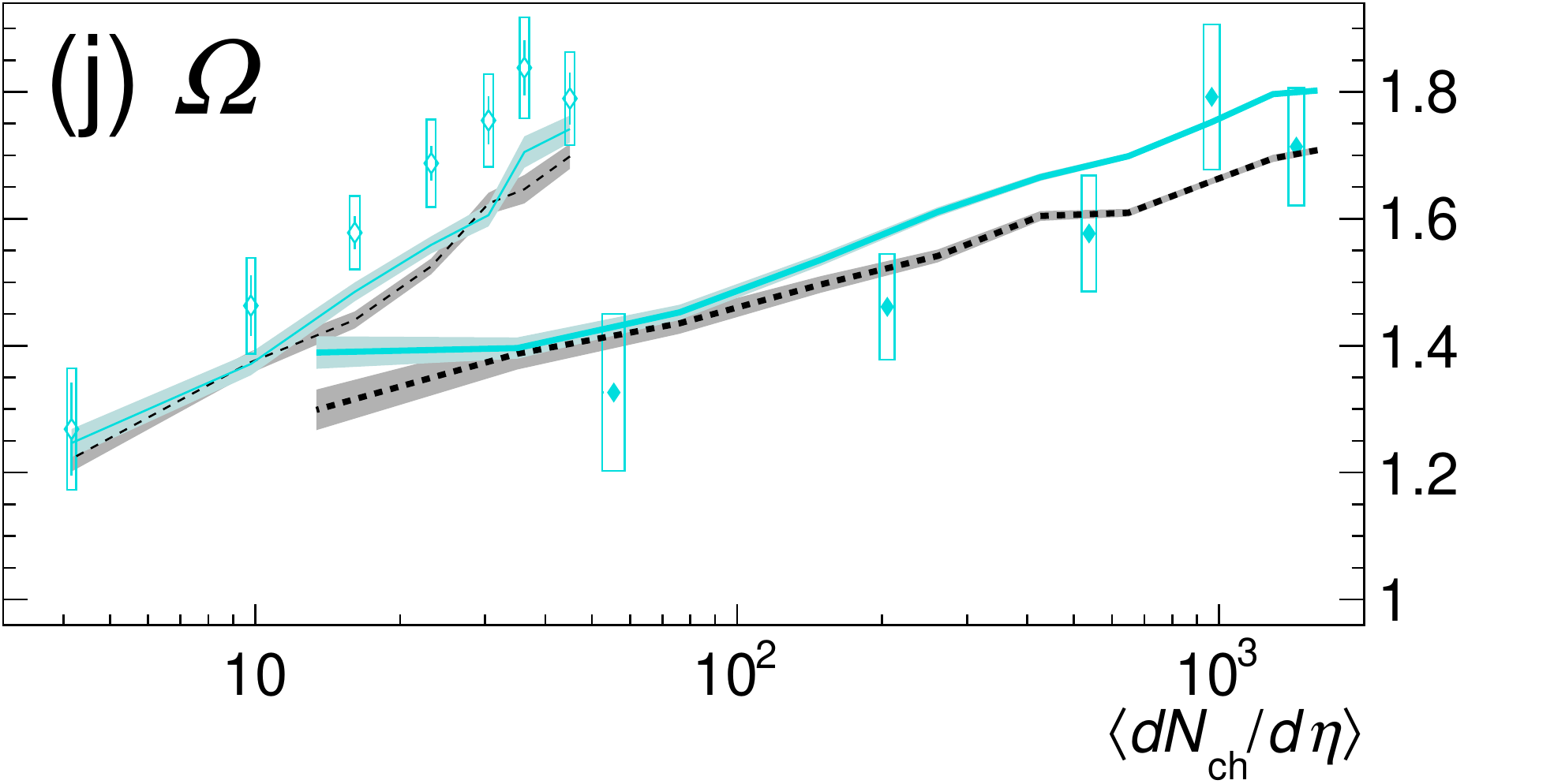}
\caption{Mean transverse momenta \mpt of different particle species in \ppb collisions at \rsnn[5.02~TeV] and \pb collisions at \rsnn (thin and thick lines, respectively). EPOS3 calculations were performed with and without a hadronic cascade modeled with UrQMD (solid and dashed lines, respectively). Values of \mpt derived from measurements by the ALICE Collaboration are also shown for \ppb~\cite{ALICE_LF_pPb,ALICE_Kstar_phi_pPb,ALICE_multistrange_pPb} and \pb collisions~\cite{ALICE_K0S_Lambda_PbPb,ALICE_Kstar_phi_PbPb,ALICE_Kstar_phi_highpT_PbPb,ALICE_multistrange_PbPb,ALICE_piKp_PbPb,ALICE_rho_PbPb} (open and filled symbols, respectively).}
\label{fig_mpt14}
\end{figure*}

\begin{figure*}[!ht]
\centering
\includegraphics[angle=0,scale=0.84]{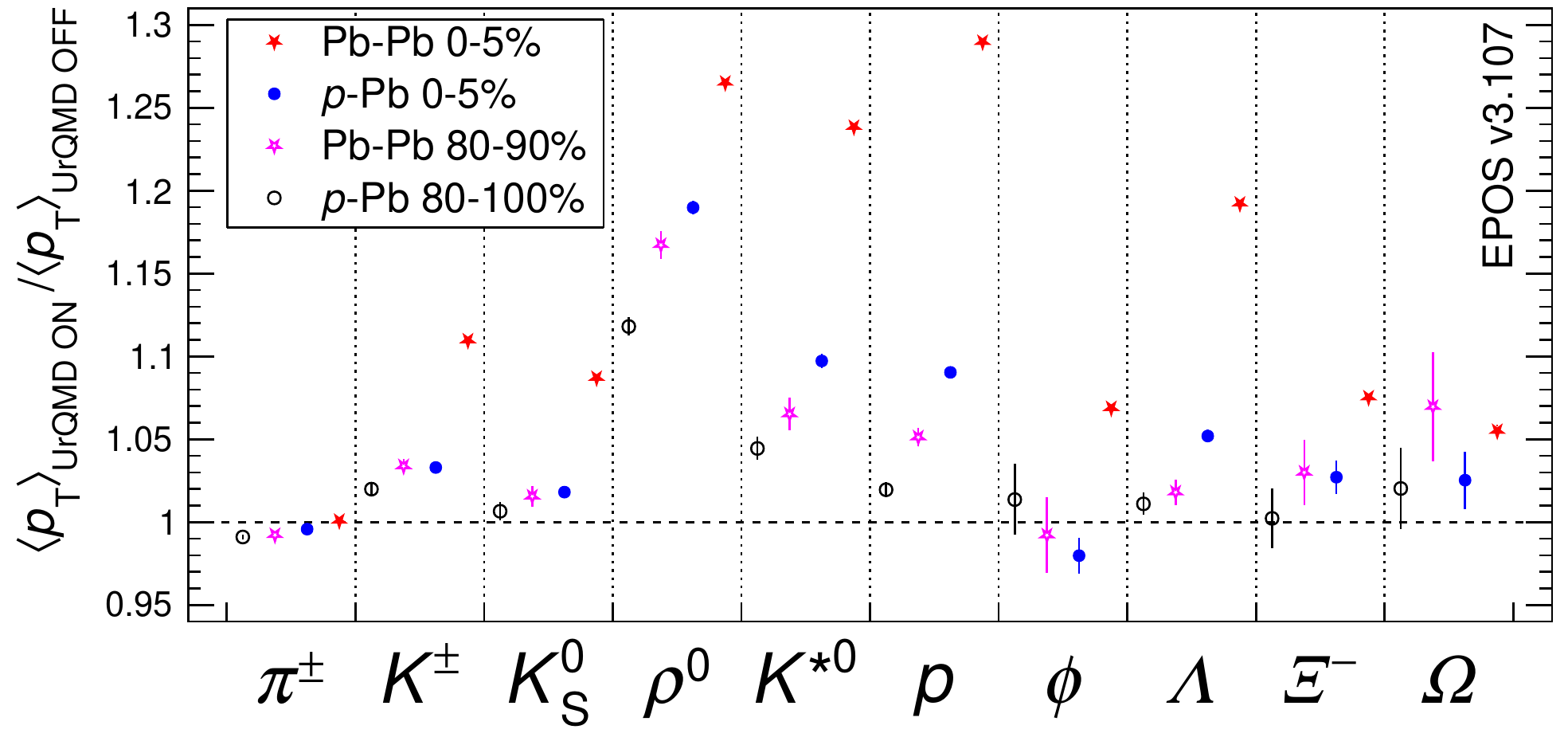}
\caption{The fractional modification of \mpt values that occurs when UrQMD is turned on, \textit{i.e.}, the ratio of \mpt values with UrQMD on to results without UrQMD. This ratio is calculated four times for each particle in different multiplicity/centrality classes: for low-multiplicity \ppb (80-100\%, black), peripheral \pb (90-100\% magenta), high-multiplicity \ppb (0-5\%, blue), and central \pb (0-5\%, red).}
\label{fig_mpt_UrQMDX}
\end{figure*}

\section{Mean Transverse Momenta}

The mean transverse momenta \mpt of various common light-flavor hadron species, including several resonances, are shown in Fig.~\ref{fig_mpt14} for \ppb collisions at \rsnn[5.02~TeV] and \pb collisions at \rsnn~\cite{Knospe_EPOS_PbPb}. The results of EPOS3 calculations with and without a hadronic cascade are compared to measurements from the ALICE Collaboration~\cite{ALICE_LF_pPb,ALICE_Kstar_phi_pPb,ALICE_Sigmastar_Xistar_pPb,ALICE_Lambdastar_pPb,ALICE_multistrange_pPb,ALICE_K0S_Lambda_PbPb,ALICE_Kstar_phi_PbPb,ALICE_Kstar_phi_highpT_PbPb,ALICE_multistrange_PbPb,ALICE_piKp_PbPb,ALICE_Lambdastar_PbPb,ALICE_rho_PbPb}, when available. For $K^{0}_{S}$, $\Lambda$, $\Xi^{-}$, and $\varOmega$ in \pb collisions, we have found the \mpt values by fitting the published ALICE \pT spectra to extrapolate the yields at low \pT. EPOS3 provides a good description of most of the measured \mpt trends in \pb collisions, but has a tendency to slightly underestimate the \mpt values for \ppb collisions. Turning on the hadronic cascade (\textit{i.e.}, turning on UrQMD) frequently increases the \mpt values and generally results in a better description of the measured data. The effect of turning on UrQMD is quantified in Fig.~\ref{fig_mpt_UrQMDX}, which shows the ratio of \mpt values with and without UrQMD for low- and high-multiplicity \ppb collisions and for peripheral and central \pb collisions for each particle species. UrQMD produces the largest changes for the short-lived resonances (\rh and \ks), as well as the proton and $\Lambda$. The effect of the hadronic cascade tends to be the greatest in central \pb collisions and decreases for smaller collision systems (consistent with the decreasing hadronic phase lifetimes discussed above). Interestingly, the short-lived \rh meson is affected even in low-multiplicity \ppb collisions, although it should be noted that even in that multiplicity class (80-100\%), the estimated hadronic phase lifetime is still the same order of magnitude as the lifetime of the \rh meson.

\section{Nuclear Modification Factors}

\begin{figure*}[!ht]
\centering
\includegraphics[angle=0,scale=0.42]{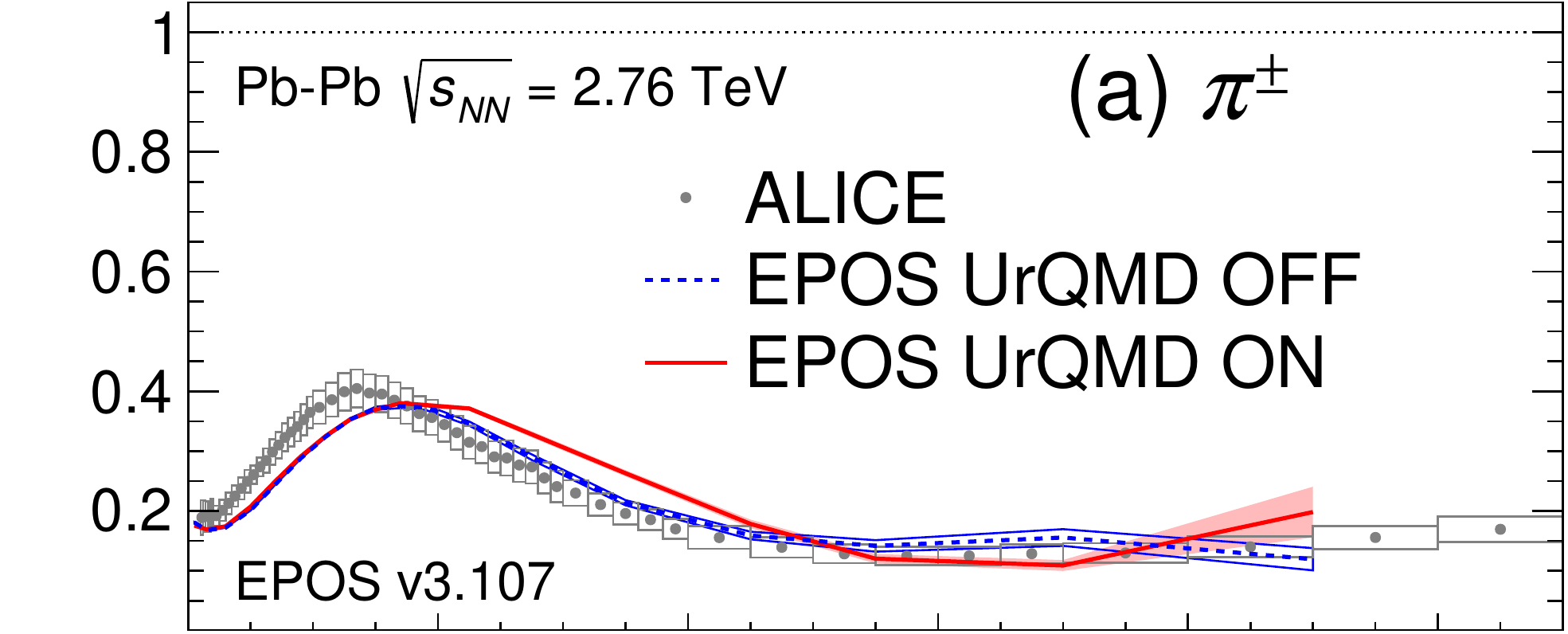}
\includegraphics[angle=0,scale=0.42]{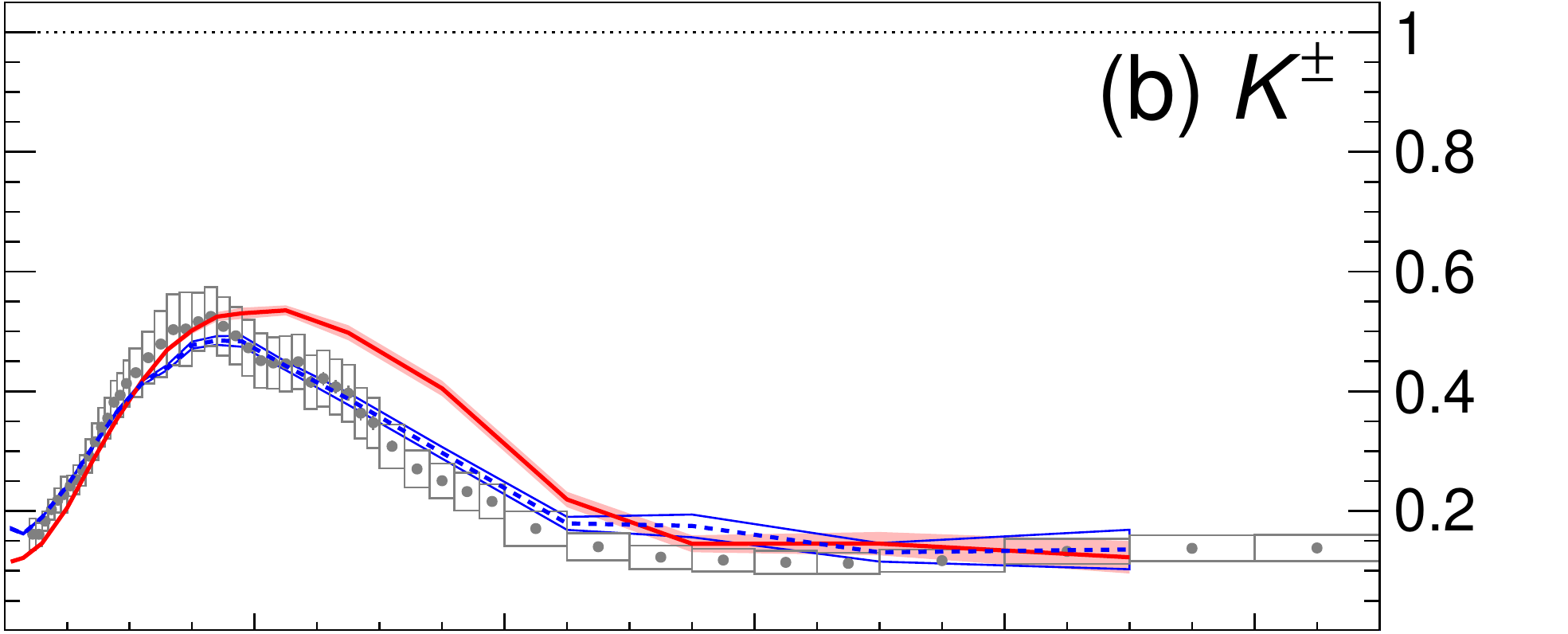}
\includegraphics[angle=0,scale=0.42]{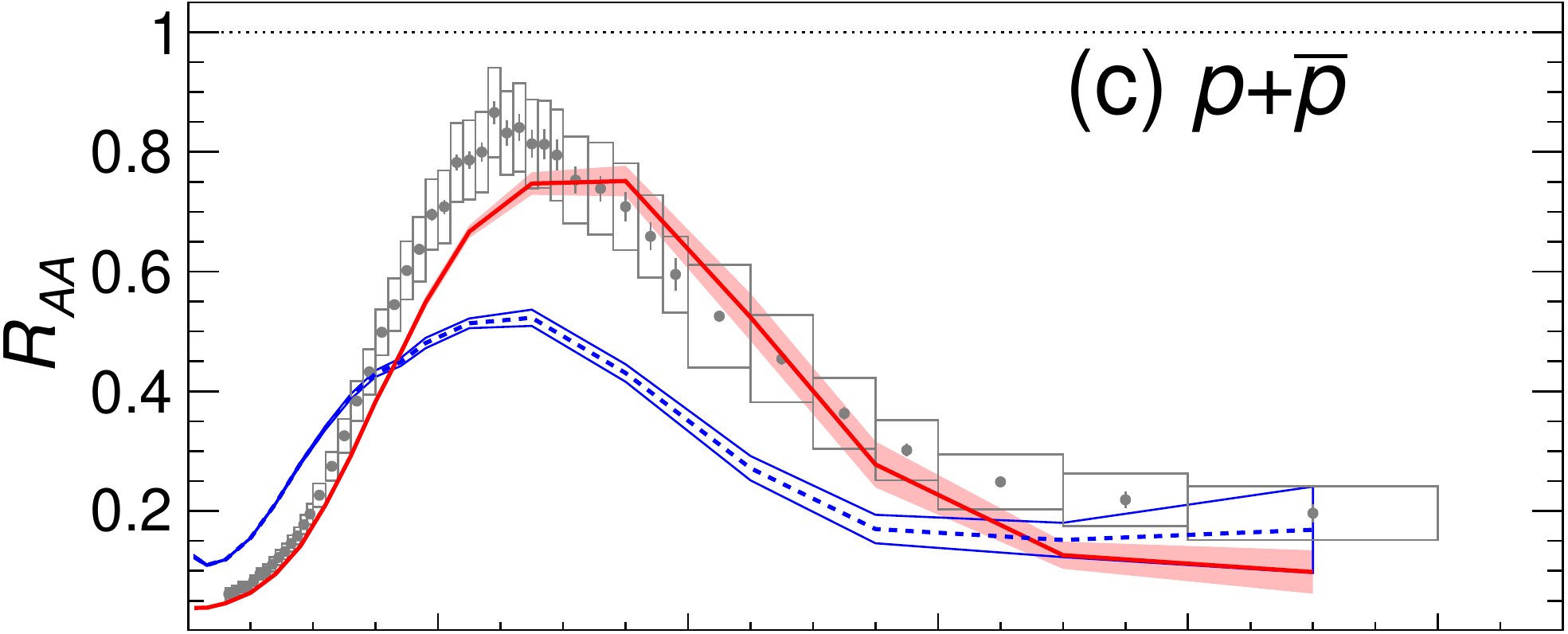}
\includegraphics[angle=0,scale=0.42]{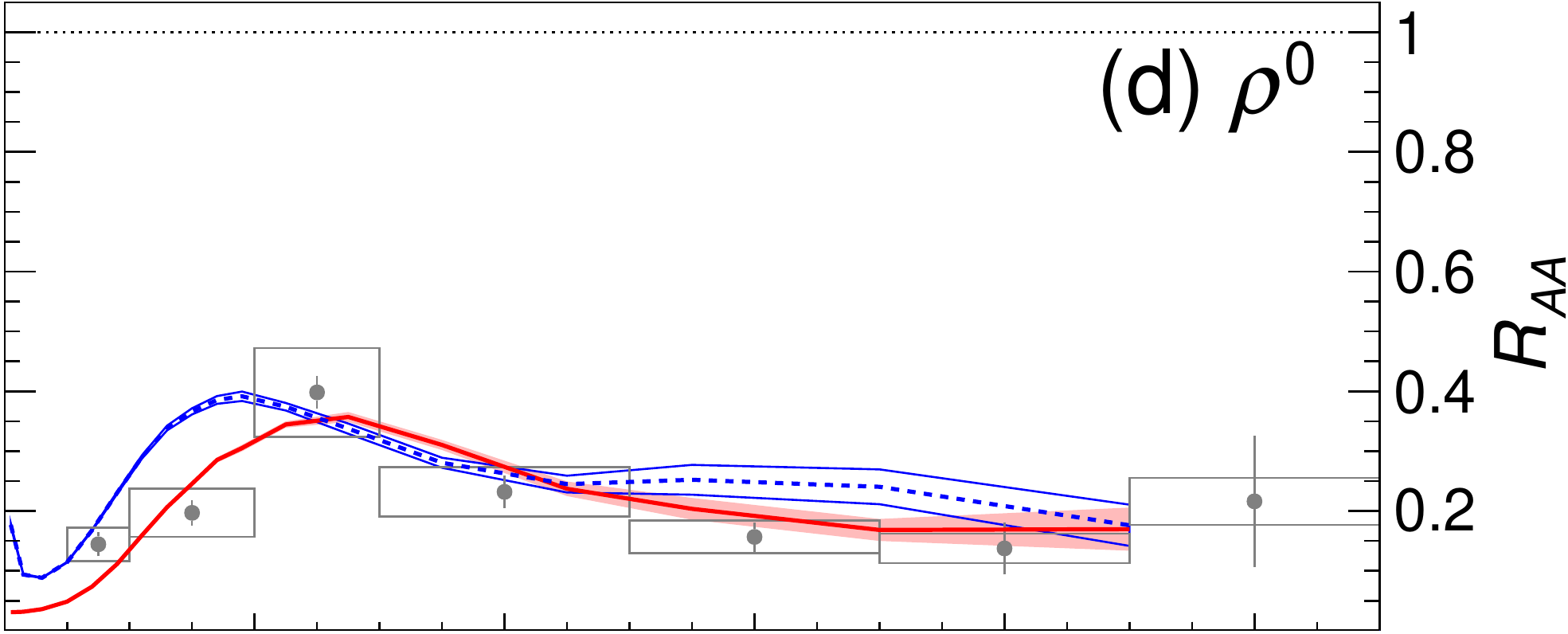}
\includegraphics[angle=0,scale=0.42]{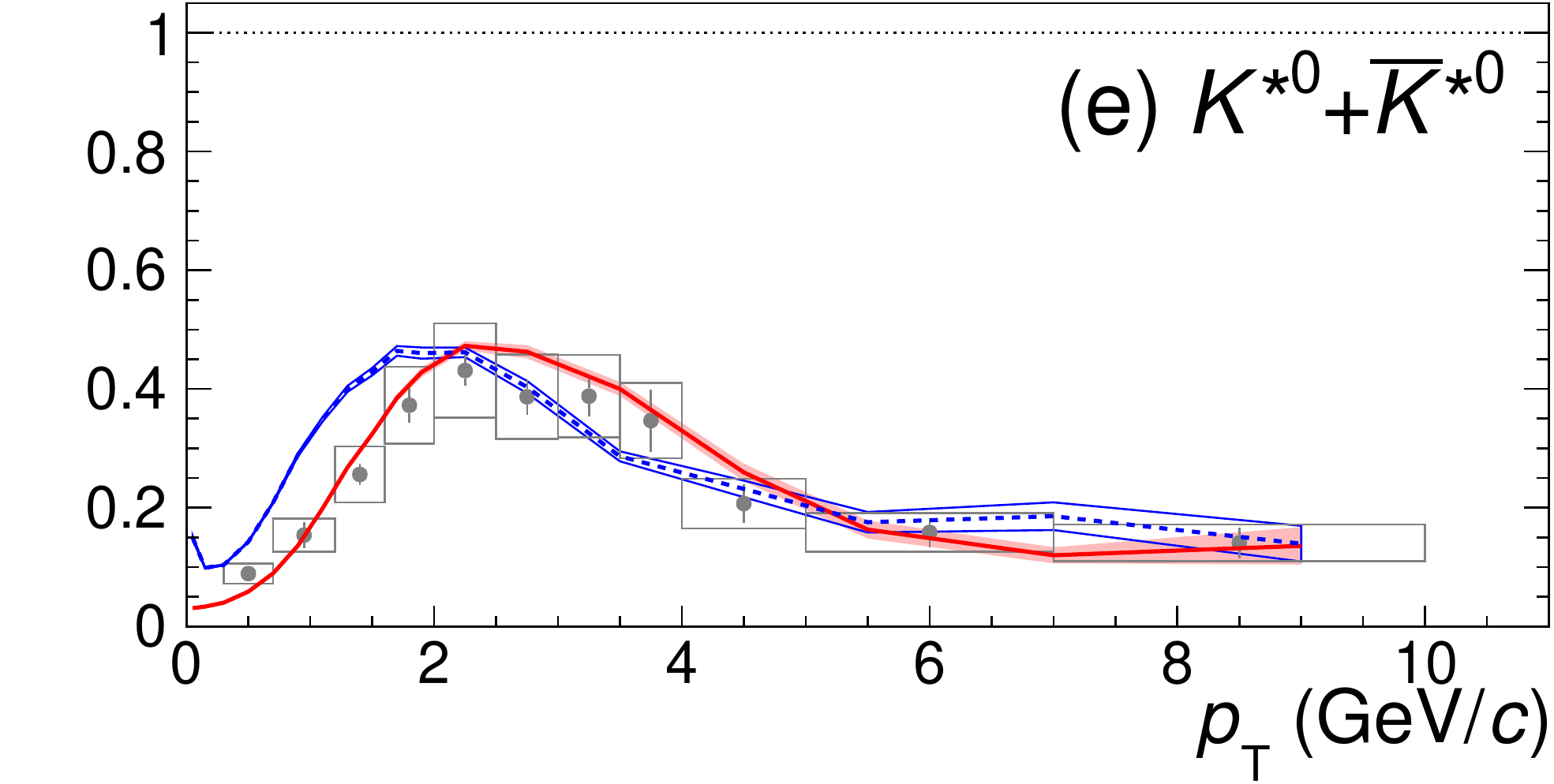}
\includegraphics[angle=0,scale=0.42]{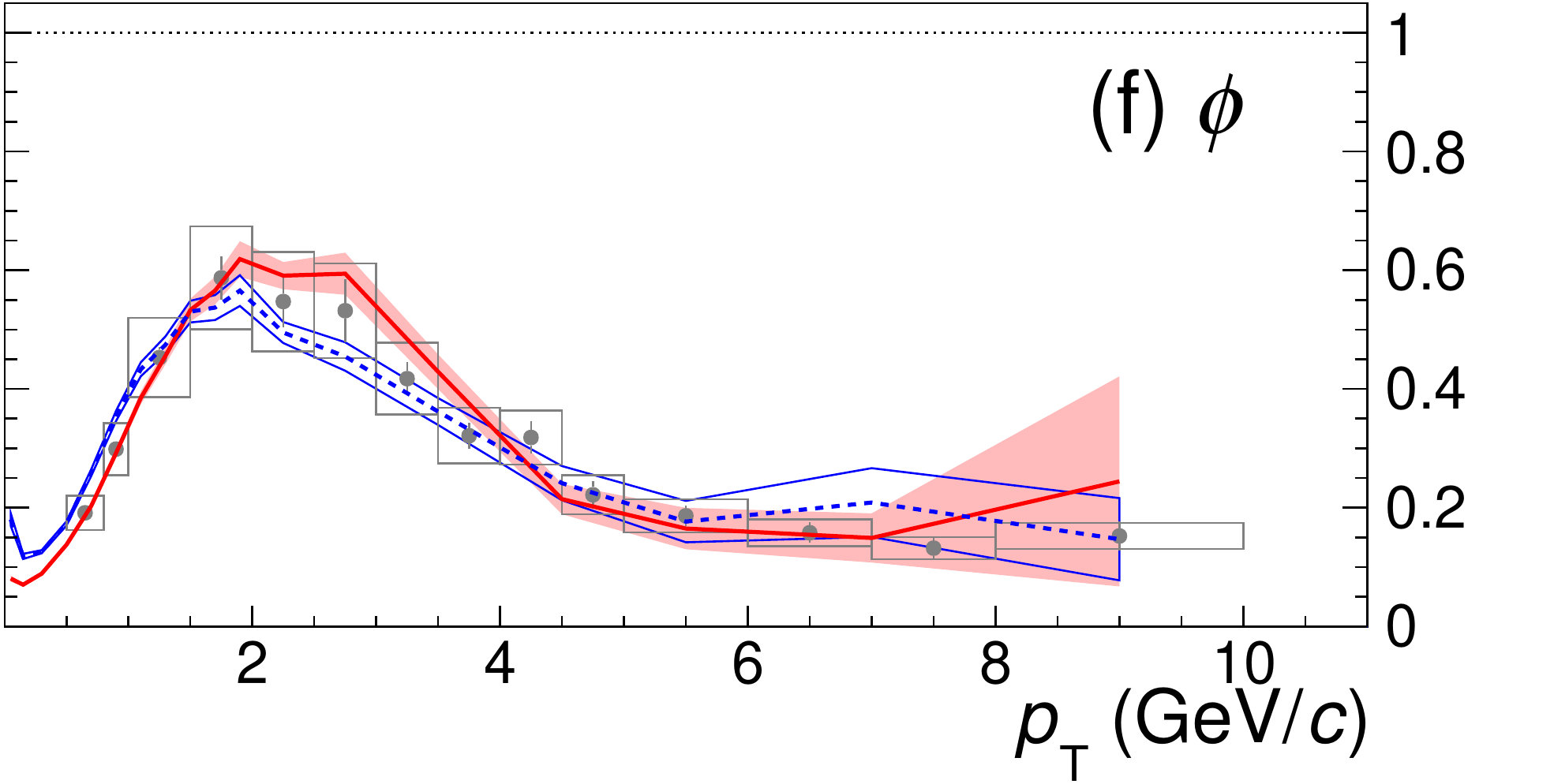}
\caption{Nuclear modification factors of $\pi^{\pm}$, $K^{\pm}$, protons, \rh, \ks, and \ph as functions of \pT in \pb collisions at \rsnn. The 0-20\% centrality class is shown for the \rh, while the 0-5\% class is shown for the other five particles. EPOS3 calculations with and without UrQMD (red and blue lines, respectively) are compared to measurements from the ALICE Collaboration~\cite{ALICE_Kstar_phi_highpT_PbPb,ALICE_piKp_PbPb,ALICE_rho_PbPb}.}
\label{fig_RAA}
\end{figure*}

EPOS3 calculations of the nuclear modification factor \raa in central \pb collisions at \rsnn are shown in Fig.~\ref{fig_RAA} and compared to measurements by the ALICE Collaboration~\cite{ALICE_Kstar_phi_highpT_PbPb,ALICE_piKp_PbPb,ALICE_rho_PbPb}. To obtain the \pp reference for these results, we generated 500 thousand \pp collisions at \rs[2.76~TeV] with UrQMD turned on, plus the same number of collisions with UrQMD turned off. The values of $\langle N_{\mathrm{coll}}\rangle$, the number of binary nucleon-nucleon collisions for each centrality class, were taken from the ALICE collaboration's Glauber-model calculation~\cite{ALICE_centrality_PbPb}. For $\pi^{\pm}$ and $K^{\pm}$, EPOS3 accurately describes the vales of \raa for high \pT; the maximum value of \raa is also well described, although it occurs at higher \pT than in the measured data. Turning on UrQMD greatly improves the description of the measured proton \raa. For the three resonances, \rh, \ks, and \ph, the measured nuclear modification factors are well described by EPOS3 with UrQMD. In contrast, removing the hadronic cascade results in a worse description for the \ks at low \pT, indicating that scattering effects do indeed modify the yields of this resonance at low \pT. Similarly, the description of \raa of the \rh meson is greatly improved in the range $1\leq\pT\leq2$~\gvc when UrQMD is turned on.

\section{Conclusions}

Our previous study of resonance production and modification with the EPOS3 model in \pb collisions has now been extended to the \ppb collision system and we have also reported new results (\textit{e.g.}, nuclear modification factors) for \pb collisions. While EPOS3 tends to overestimate the yield ratios of resonances to ground-state particles, it is able to qualitatively describe the evolution of most of those ratios with system size across \ppb and \pb collisions. While shorter-lived resonances tend to be suppressed in the larger collision systems, our calculations predict that the \ssl and \dlp ratios should not be suppressed even in central \pb collisions. This prediction may be testable in future measurements at the LHC. Our calculated \ksk and \rhpi ratios are suppressed in high-multiplicity \ppb collisions; there are hints of a similar trend in measurements from the ALICE Collaboration. The EPOS3 \pT spectra for resonances tend to agree best with the measured data for intermediate \pT ($2\lesssim\pT\lesssim4$~\gvc) and when UrQMD, which models interactions in the hadronic phase, is turned on. Turning on UrQMD also tends to improve the description of the mean transverse momentum values of many species of light-flavor hadrons and results in large increases in \mpt for short-lived resonances, protons, and $\Lambda$. EPOS3 is also able to reproduce the nuclear modification factors of resonances in central \pb collisions at \rsnn. These results highlight the importance of the hadronic phase, with an estimated lifetime $\sim$1~fm/$c$ in \ppb collisions and 0.5--10~fm/$c$ in \pb collisions, in determining the final resonance yields and \pT distributions. Additional effects arise from the interplay between particle production in the core and corona parts of the collision. The effects of the hadronic phase are found to be most important at low \pT ($\lesssim1$-2~\gvc) in both \ppb and \pb collisions, which modifies the \pT distributions of these resonances and their decay products. Modifications of the \pT distributions, including re-scattering, regeneration, annihilation, and radial flow, will also affect hadron correlation measurement. A rigorous description of the hadronic phase, even in small collision systems such as \ppb, is therefore essential for a complete understanding of many different observables in studies of ion-ion collisions.

\begin{acknowledgments}

The authors would like to thank J. Aichelin for his fruitful discussions and valuable contributions. AK thanks R. Bellwied for his support of this work. This work was supported by the U.S. Department of Energy Office of Science under contract numbers DE-FG02-07ER41521 and DE-SC0013391. This work was supported by GSI and the Hessian initiative for excellence (LOEWE) through the Helmholtz International Center for FAIR (HIC for FAIR). The authors acknowledge the Texas Advanced Computing Center (TACC) at the University of Texas at Austin for providing computing resources that have contributed to the research results reported within this paper. URL: http://www.tacc.utexas.edu. Additional computational resources were provided by Le Centre de Calcul de l'IN2P3 and the LOEWE Frankfurt Center for Scientific Computing (LOEWE-CSC). JS thanks the BMBF through the ErUM-Data project for funding.

\end{acknowledgments}

\end{document}